%% file: AI_RAN_economy.tex
\documentclass[conference, nofonttune]{IEEEtran}
\IEEEoverridecommandlockouts
\usepackage{amsmath,amsfonts}
\usepackage{algorithmic}
\usepackage{algorithm}
\usepackage{array}
\usepackage[caption=false,font=scriptsize,labelfont=sf,textfont=sf]{subfig}
\usepackage{textcomp}
\usepackage{stfloats}
\usepackage{url}

\usepackage{verbatim}
\usepackage{graphicx}
\usepackage{xcolor}
\usepackage{cite}
\usepackage{booktabs}
\usepackage{pgfplots}
\usepackage{pgfplotstable}
\usepackage{tikz}
\usetikzlibrary{patterns}
\usepackage{lipsum}  
\pgfplotsset{compat=1.18}
\usepgfplotslibrary{groupplots}
\usepgfplotslibrary{statistics} % for boxplots
\usepackage[acronym,nomain]{glossaries}
\usepackage{soul}
\input{acronyms.tex}

\begin{document}
\bstctlcite{IEEEexample:BSTcontrol}

%\title{Modeling AI-RAN Economics: Toward Revenues with Infrastructure Sharing} % or something else, but a bit more assertive than the current
%\title{Rethinking RAN Economics: Modeling Cost and Revenue Opportunities Towards AI-RAN}

% \title{Cost Modeling and Revenue Opportunities in Open and Programmable Architectures: A Techno-Economic Framework for AI-RAN} %TM

% \title{Revenue Opportunities and Cost Modeling for\\Open and Programmable AI-RAN:\\A Techno-Economic Framework} %TM

\title{A Techno-Economic Framework for Cost Modeling and Revenue Opportunities in Open and Programmable AI-RAN}

\author{Gabriele Gemmi, Michele Polese, Tommaso Melodia\\
Institute for Intelligent Networked Systems, Northeastern University\\
\{g.gemmi, m.polese, melodia\}@northeastern.edu
\vspace{-0.3cm}
\thanks{This work is partially supported by the U.S. NSF under award TI-2449452 and by OUSD(R\&E) through Army Research Laboratory Cooperative Agreement Number W911NF-24-2-0065. The views and conclusions contained in this document are those of the authors and should not be interpreted as representing the official policies, either expressed or implied, of the Army Research Laboratory or the U.S. Government. The U.S. Government is authorized to reproduce and distribute reprints for Government purposes notwithstanding any copyright notation herein.}}

\maketitle

\glsunset{airan}

\begin{abstract}
The large-scale deployment of 5G networks has not delivered the expected return on investment for mobile network operators, raising concerns about the economic viability of future 6G rollouts. At the same time, surging demand for \gls{ai} inference and training workloads is straining global compute capacity. AI-RAN architectures, in which \gls{ran} platforms accelerated on \gls{gpu} share idle capacity with AI workloads during off-peak periods, offer a potential path to improved capital efficiency. However, the economic case for such systems remains unsubstantiated. In this paper, we present a techno-economic analysis of AI-RAN deployments by combining publicly available benchmarks of 5G Layer-1 processing on heterogeneous platforms---from x86 servers with accelerators for channel coding to modern \glspl{gpu}---with realistic traffic models and AI service demand profiles for \gls{llm} inference. We construct a joint cost and revenue model that quantifies the surplus compute capacity available in \gls{gpu}-based RAN deployments and evaluates the returns from leasing it to AI tenants.  Our results show that, across a range of scenarios encompassing token depreciation, varying demand dynamics, and diverse GPU serving densities, the additional capital and operational expenditures of \gls{gpu}-heavy deployments are offset by AI-on-RAN revenue, yielding a return on investment of up to $\mathbf{8\times}$. These findings strengthen the long-term economic case for accelerator-based RAN architectures and future 6G deployments.
\end{abstract}


\begin{picture}(0,0)(25,-460)
    \put(0,0){
    \put(0,0){\footnotesize \scshape This paper has been accepted for publication on the 35th International Conference on Computer Communications and Networks (ICCCN 2026).}
     \put(0,-10){
     \scriptsize\scshape \textcopyright~2026 IEEE. Personal use of this material is permitted. Permission from IEEE must be obtained for all other uses, in any current or future media, including}
     \put(0, -17){
     \scriptsize\scshape reprinting/republishing this material for advertising or promotional purposes, creating new collective works, for resale or redistribution to servers or}
     \put(0, -24){
     \scriptsize\scshape lists, or reuse of any copyrighted component of this work in other works.}
     }
 \end{picture}

\glsresetall
\glsunset{airan}
\glsunset{oran}

% \fi
\IEEEpeerreviewmaketitle

\setlength{\textfloatsep}{8pt}
\section{Introduction}
Over the past decade, mobile network operators have completed large-scale \gls{5g} deployments, upgrading spectrum holdings, transport networks, and radio access infrastructure.
Despite these investments, the expected revenue uplift has not materialized.
Most operators have experienced flat or declining returns, and no transformative application has emerged to justify the capital expenditure profile of the \gls{5g} rollout~\cite{gsma2023globaltrends}.
As the community begins to define \gls{6g} system requirements, this economic misalignment highlights a structural challenge: without a new source of value creation, the incentive to undertake a further hardware refresh remains weak.

In parallel, \gls{ai} has become a dominant driver of global compute demand.
Training and inference workloads increasingly rely on high-performance accelerators, creating sustained pressure on datacenter capacity~\cite{mckinsey2023genai}.
These workloads also exhibit significant spatial and temporal variability, suggesting potential complementarity with the utilization profile of the \gls{ran} \cite{BurstGPT}.
In virtualized and \gls{oran} architectures, where Layer-1 processing increasingly relies on general-purpose accelerators such as \glspl{gpu}, the same hardware supporting the \gls{ran} may be repurposed for different compute tasks when traffic intensity is low, with utilization gains driven by statistical multiplexing.

This concept, often referred to as \emph{AI-and-RAN} by the AI-RAN Alliance, envisions the \gls{ran} not only as a consumer of \gls{ai} algorithms but also as a provider of compute resources to external tenants.
Under this model, idle accelerator capacity during off-peak periods can be allocated to \gls{ai} tasks (also known as \emph{AI-on-RAN} workloads), thereby introducing a new revenue mechanism while preserving the real-time performance constraints of the radio workload.
If realized at scale, such architectures may help address the economic feasibility of future \gls{6g} deployments by improving the capital efficiency of accelerator-rich \gls{ran} platforms. However, despite research and technical developments in this space~\cite{polese2025beyond,airan_alliance2024,kundu2025airan,shah2025caora,shah2025proactive}, there are still several questions around the economics of such systems. Among others: will the revenue generated by AI-on-RAN offset the increased \gls{capex} associated with powerful general purpose accelerators? \emph{Are systems dynamics (e.g., utilization patterns over time) amenable to generating enough revenue through sharing?} 

In this paper, we answer these questions by investigating the techno-economic implications of \gls{airan} systems.
We combine publicly available benchmarks of \gls{5g} Layer-1 processing on heterogeneous platforms, including x86 servers with \gls{fec} accelerators to modern \glspl{gpu}, with user-driven traffic models to quantify the amount of surplus compute capacity expected in realistic deployments. We also model demand for \gls{ai} services, namely, \glspl{llm}, and combine the models to construct a cost and revenue model that characterizes the value of renting \gls{ran} excess capacity to \gls{ai} tenants. 
% \hl{Something more quantitative here. What's the answer to the questions?}
Our results indicate that in most scenarios modeled in this paper---including token depreciation, varying demand dynamics, and diverse serving densities for LLMs on GPUs---the additional \gls{capex} and \gls{opex} introduced by \gls{gpu}-heavy deployments are offset through revenue generated by AI-on-RAN services running on the shared AI-RAN infrastructure, with a return on investment of up to 8 times. This strengthens
% Our results illustrate conditions under which operators can generate meaningful ancillary revenue from unused \gls{gpu} resources, thereby strengthening 
the long-term economic case for accelerator-based \gls{ran} architectures and future \gls{6g} rollouts. We also release the software used for the techno-economic model as open source and also packaged as a webapp that can be used to reproduce the results of this paper.\footnote{Repo: www.github.com/wineslab/AIRAN-revenue-model}\footnote{Webapp: www.open6g.us/\#/ai-ran-economics}

% The remainder of this article is organized as follows.
% Section~\ref{sec:ran_cost} introduces the \gls{ran} architecture landscape, defines a unified baseband capacity metric, and compares the cost and power efficiency of \gls{gpu}-based and \gls{fec}-accelerated server platforms.
% Section~\ref{sec:ran_demand} develops a demand-driven deployment model that sizes the server cluster for a given area and population density, and characterizes the time-varying residual capacity available for non-\gls{ran} workloads.
% Section~\ref{sec:llm_demand} formulates the \gls{llm} inference demand model that consumes this surplus compute.
% Section~\ref{sec:roi} combines the deployment cost and inference revenue into a unified economic framework.
% Section~\ref{sec:scenarios} instantiates the models for a dense urban deployment in Milan, Italy, and presents quantitative results.
% Finally, Section~\ref{sec:conclusion} summarizes the findings and discusses future directions.
The remainder of this article is organized as follows. Section~\ref{sec:related} surveys related work, Section~\ref{sec:ran_cost} introduces the \gls{ran} architecture landscape and compares \gls{gpu}-based and \gls{fec}-accelerated platforms. Section~\ref{sec:ran_demand} sizes the server cluster for a target deployment and characterizes residual capacity available for non-\gls{ran} workloads. Section~\ref{sec:llm_demand} formulates the \gls{llm} inference demand model consuming this surplus. Section~\ref{sec:roi} combines deployment cost and inference revenue into a unified economic framework. Section~\ref{sec:scenarios} presents quantitative results for a dense urban deployment in Milan, Italy. Finally, Section~\ref{sec:conclusion} summarizes findings and discusses future directions.
\section{Related Work}\label{sec:related}

The economic case for virtualized and open RAN has been the subject of extensive industry analysis. Analysys Mason concluded that Open RAN can deliver TCO savings of up to 30\% under favorable conditions~\cite{analysys_mason2022}, and a follow-up study reported growing operator confidence, with Open RAN projected to reach 20--30\% of total RAN revenues by 2028~\cite{analysys_mason2024,delloro2024}. These analyses, however, evaluate the RAN compute substrate as single-purpose infrastructure and do not consider the additional revenue that idle accelerator capacity could generate.

The AI-RAN Alliance, formed in 2024~\cite{airan_alliance2024}, introduced a framework for repurposing that idle capacity by organizing research into three pillars---AI-for-RAN, AI-on-RAN, and AI-and-RAN---with the latter targeting the coexistence of RAN and AI workloads on shared GPU hardware. 
% Early architectural work by Kelkar and Dick~\cite{kelkar2021aerial} demonstrated concurrent 5G and AI processing on NVIDIA Aerial, and 
Kundu~et~al.~\cite{kundu2025airan} proposed a reference architecture with a proof-of-concept on GH200 servers, establishing feasibility without addressing economics. On the orchestration side, Shah~et~al.~\cite{shah2025caora,shah2025proactive} developed the CAORA framework, which uses reinforcement learning to dynamically partition MIG resources while maintaining near-99\% RAN demand fulfillment, and Polese~et~al.~\cite{polese2025beyond} extended the O-RAN \gls{smo} to support unified management of heterogeneous workloads. These contributions address the \emph{how} of AI-RAN coexistence but leave the \emph{whether it pays off} question unanswered.

On the demand side, Erdil~\cite{erdil2025inference} developed a roofline-style cost model for \gls{llm} serving that relates token throughput to hardware constraints, while Gundlach~et~al.~\cite{gundlach2025price} empirically documented quality-adjusted inference price declines of $5$--$10\times$ per year. Demirer~et~al.~\cite{demirer2025market} corroborated these trends using API transaction data from OpenRouter and Microsoft Azure, reporting approximately $1000\times$ price declines for models at the 2023 frontier and estimating short-run price elasticities just above unity. From a model efficiency standpoint, Xiao~et~al.~\cite{xiao2025densing} introduced the concept of \emph{capability density} and showed that the maximum capability density of open-source LLMs doubles roughly every 3.5~months, implying that inference cost per unit of capability decreases exponentially---a trend directly relevant to our $\rho_\mathrm{tok}$ parameter. The BurstGPT dataset~\cite{wang2024burstgpt}, which we use to derive weekly demand profiles, provides timestamped traces of real-world \gls{llm} inference traffic. 

Although edge \gls{llm} deployment has received attention from a systems efficiency perspective~\cite{cai2025edgellm,ding2024hybrid}, no prior work treats the RAN as a source of opportunistic inference compute or models the resulting revenue against the GPU cost premium. This paper fills that gap with a unified techno-economic framework combining platform benchmarking, demand-driven deployment sizing, and a dual-use revenue analysis.

\section{RAN Architectures and Cost Model}
\label{sec:ran_cost}

\begin{figure}[t]
  \centering
  \includegraphics[width=\linewidth]{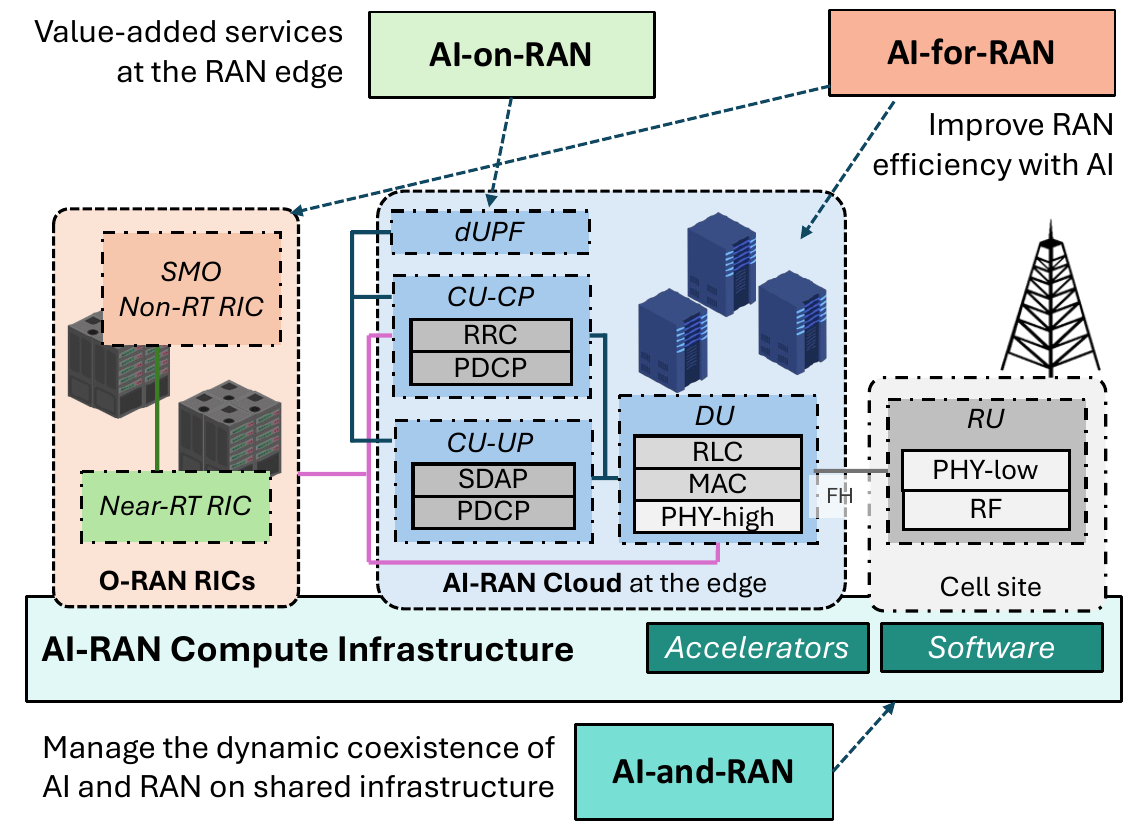}
  % \rule{\textwidth}{0.4pt}
  % \vspace{4cm}
  % \rule{\textwidth}{0.4pt}
    \setlength{\abovecaptionskip}{-.1cm}
  \caption{\Gls{airan} system architecture and role of AI-for-RAN (increasing RAN efficiency and performance with AI), AI-on-RAN (value-added edge services co-deployed with the RAN and with access to RAN data and telemetry), and AI-and-RAN (orchestration and management to support the coexistence of RAN, AI-for-RAN, and AI-on-RAN).}
  \label{fig:architecture}
\end{figure}

Traditional \glspl{ran} rely on integrated, vendor-specific base stations in which radio, baseband processing, and control functions are tightly coupled within proprietary hardware.
The 3GPP and \gls{oran} architectures in 5G and beyond disaggregate these functions into distinct logical components: the \gls{ru} handles radio frequency operations, the \gls{du} processes Layer-1 and Layer-2 protocol functions, and the \gls{cu} manages higher-layer control and user plane processing (Fig.~\ref{fig:architecture}). While the \gls{ru} is usually based on \glspl{fpga} or dedicated hardware, the \gls{du} and \gls{cu} open opportunities for software-based implementations deployed on general purpose compute~\cite{kelkar2021aerial,garcia2021ran}. 
% By standardizing interfaces between these components and virtualizing baseband processing on general-purpose compute platforms, \gls{oran} enables multi-vendor interoperability and offers operators greater flexibility.

\Gls{airan} extends this architecture by explicitly designing the compute substrate to support both real-time Layer-1 processing and batch or inference AI-on-RAN workloads from external \gls{ai} tenants.
The key enabler is the use of high-performance accelerators, such as \glspl{gpu}, whose resources can be dynamically allocated to \gls{ran} functions (e.g., during peak traffic periods) and to \gls{ai} tasks (e.g., during idle intervals).
The economic viability of this dual-use model depends critically on the baseband capacity, cost, and power efficiency of the underlying hardware platforms, which we characterize below.

\subsection{Baseband Capacity Metric}

\input{server_table.tex}

To compare platforms with heterogeneous \gls{mimo} layers, cell count, and bandwidth configurations under a single metric, we define the baseband capacity $B$ [MHz] as the aggregate downlink baseband processing capacity:
\begin{equation}
\label{eq:mhzlayers}
B = L_{\text{DL}} \times N_C \times \text{BW},
\end{equation}
where $L_{\text{DL}}$ is the number of downlink \gls{mimo} layers, $N_C$ is the number of simultaneously served cells, and $\text{BW}$ is the channel bandwidth per cell in MHz.
This product captures the total spatial-frequency processing load sustained by a baseband processor (e.g., a server) and provides the basis for two economic efficiency indicators.
Capital efficiency relates baseband capacity to the server acquisition cost:
$\eta_C = \frac{B}{\text{Cost}}$ while power efficiency relates it to the platform power consumption: $\eta_O = \frac{B}{\text{Power}}$
Therefore, the first represent a component of \gls{capex}, while the second contributes to \gls{opex}.

\subsection{Platform Comparison}

Table~\ref{tab:capacity_summary} summarizes the baseband processing capacity and economic efficiency of representative server platforms from publicly available benchmarks and vendor specifications.
The platforms are grouped by Layer-1 software stack: NVIDIA Aerial~\cite{kelkar2021nvidia}, which targets \gls{gpu}-based architectures using the GH200 Grace Hopper Superchip, and Intel FlexRAN~\cite{wang2024flexran}, which leverages the VRB1 \gls{fec} accelerator~\cite{intelvrb2} on x86 server platforms. The literature and technical specifications for both platforms~\cite{kelkar2021nvidia,wang2024flexran,intelvrb2} identifies two representative cell configurations as benchmarks: a \emph{macro-cell} with up to 16~\gls{mimo} layers transmitted through a mMIMO frontend, and a \emph{micro-cell} with 4~\gls{mimo} layers transmitted through a low-order antenna panel.

The \gls{gpu}-based ARS-111GL achieves 9600~MHz in the macro-cell configuration and up to 16~GHz in micro-cell mode, both at a platform cost of \$45000 and 1200~W power draw.
The VRB1-based FlexRAN platforms (EGX74I at \$6000 and DL110 at \$7200, both at 300~W) reach 9600~MHz in the macro-cell configuration---matching the \gls{gpu} in raw baseband throughput---at a fraction of the cost and power.
As discussed in~\cite{wang2024flexran,intelvrb2}, for the micro-cell mode reference solution, capacity drops to 1440~MHz with correspondingly lower efficiency.

In a realistic dense urban deployment, macro and micro cells coexist.
Following ITU-R deployment guidelines~\cite{itu2412}, we adopt a 1:3 macro-to-micro cell ratio for the remainder of the analysis.
The deployment-weighted average baseband capacity per server is:
\begin{equation}
\label{eq:B_avg}
\overline{B} = \frac{B_\mathrm{macro} + 3\,B_\mathrm{micro}}{4}.
\end{equation}
For Aerial, $\overline{B}_\mathrm{Aerial} = 14400$~MHz, while for FlexRAN, $\overline{B}_\mathrm{FlexRAN} = 3480$~MHz.
Despite Aerial's higher absolute capacity, FlexRAN achieves a superior CapEx efficiency: $\eta_C^\mathrm{FlexRAN} \approx 0.70$~MHz/\$ versus $\eta_C^\mathrm{Aerial}\approx 0.32$~MHz/\$.
This is because Aerial's sevenfold cost premium outweighs its fourfold capacity advantage in the mixed deployment, making FlexRAN the more capital-efficient baseband platform---absent the dual-use revenue considered in the following sections.

To translate these per-server metrics into deployment costs, we dimension each stack to deliver an aggregate peak throughput of 10~Gbps---assuming $\mathrm{SE} = 9$~bit/s/Hz and 20\% L1/L2 overhead---and compute the 10-year \gls{tco}, reflecting a typical mobile network investment cycle.

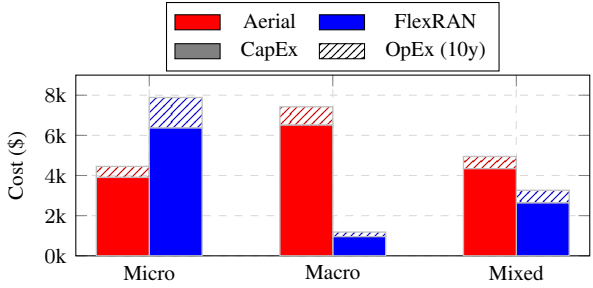
\begin{figure}[t]
  \centering
  \input{figures/tco_pgfplot}
    \setlength{\abovecaptionskip}{-.1cm}
  \caption{\gls{tco} for 10 Gbps aggregate peak throughput over 10 years.}
  \label{fig:tco_plot}
\end{figure}

Figure~\ref{fig:tco_plot} reports the resulting CapEx, OpEx, and \gls{tco} broken down by cell type and platform stack.
In the macro-cell configuration, FlexRAN achieves the lowest \gls{tco} thanks to its high per-server capacity at low cost and power.
In the micro-cell and mixed 3:1 deployments, the gap narrows as the higher cell count reduces FlexRAN's efficiency advantage.
However, the economic case for \gls{airan} rests not on baseband cost alone but on the \emph{dual-use} capability of \gls{gpu} hardware: during off-peak hours, the \gls{gpu} can be repurposed for \gls{ai} inference workloads---an option unavailable to dedicated \gls{fec} accelerators.
The higher acquisition cost of \gls{gpu} platforms must therefore be weighed against the additional revenue they can generate from surplus compute, which we quantify in the following sections.

\section{RAN Demand and Deployment Model}
\label{sec:ran_demand}

We develop a general model for sizing an \gls{airan} deployment given a target geographic area and population density, deriving the number of servers required, their associated capital and operational costs, and the residual compute capacity available for \gls{ai} workloads.
The model is anchored to the \gls{imt}-2030 (\gls{6g}) capability targets defined in \gls{itu}-R Recommendation M.2160~\cite{itu2160}, which sets a per-user experienced downlink rate of 300--500~Mbps for the immersive communication usage scenario---a $3\times$ improvement over the \gls{imt}-2020 requirement of 100~Mbps~\cite{itu2410}.
Since no outdoor dense urban area traffic capacity target has been finalized for \gls{imt}-2030 at the time of writing, we adopt a user-demand-driven approach to derive the required area capacity.

We model the per-user downlink rate $R_\mathrm{user}(w,h)$ as the product of a long-term growth trend and a normalized hourly demand profile $\lambda_\mathrm{RAN}(h) \in [0,1]$, with $\max_h \lambda_\mathrm{RAN}(h) = 1$ at the busy hour:
\begin{equation}
\label{eq:r_user_trend}
R_\mathrm{user}(w,h) = R_\mathrm{user}(0) \cdot \rho_\mathrm{R}^{w/52} \cdot \lambda_\mathrm{RAN}(h)
\quad [\text{Mbps}],
\end{equation}
where $w$ is the week index (with $w=0$ at deployment), $h \in \{0,\ldots,23\}$ is the hour of day, $\rho_\mathrm{R}$ is the annual RAN demand growth factor, and $R_\mathrm{user}(0)$ is the baseline busy-hour rate from \gls{imt}-2030 (e.g., 300~Mbps).
Given a deployment area with population density $\rho_\mathrm{pop}$, the downlink demand per unit area at week $w$, hour $h$ is:
\begin{equation}
\label{eq:area_demand}
D_\mathrm{area}(w,h) = \rho_\mathrm{pop} \cdot \eta_\mathrm{pen} \cdot \alpha_\mathrm{BH} \cdot R_\mathrm{user}(w,h)
\quad [\text{Mbps/km}^2],
\end{equation}
where $\eta_\mathrm{pen}$ is the smartphone penetration and $\alpha_\mathrm{BH}$ is the busy-hour concurrency factor.

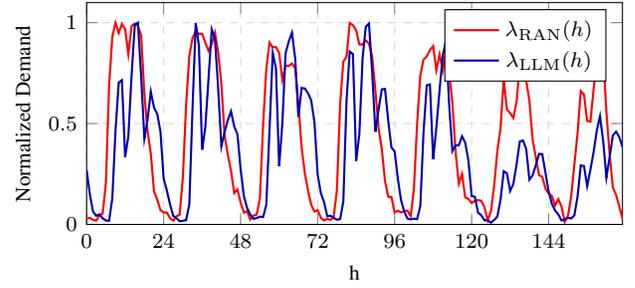
\begin{figure}[t]
  \centering
    \setlength{\abovecaptionskip}{-.1cm}
  \input{figures/daily_usage_pgfplot}
  \caption{Weekly usage patterns for \gls{ran} and \gls{llm} workloads showing slightly complementary demand cycles.}
  \label{fig:daily_usage}
\end{figure}

To translate this demand into a server count, each platform provides an average baseband capacity $\overline{B}$ (defined in Eq.~\eqref{eq:mhzlayers}), computed as the mean across the benchmark configurations of the selected Layer-1 stack (Table~\ref{tab:capacity_summary}).
After accounting for average spectral efficiency and L1/L2 signaling overhead $\eta_\mathrm{OH}$ (\gls{dmrs}, \gls{csirs}, \gls{ssb}, \gls{prach}, guard bands), the net deliverable downlink throughput per server is:
\begin{equation}
\label{eq:throughput}
C_\mathrm{net} = \overline{B} \cdot \mathrm{SE} \cdot (1 - \eta_\mathrm{OH})
\quad [\text{Mbps}].
\end{equation}
For a target deployment area $A$~[km$^2$], we denote by $G^\mathrm{RAN}(w,h)$ the number of GPUs (servers) required for RAN at week $w$, hour $h$:
\begin{equation}
\label{eq:g_RAN}
G^\mathrm{RAN}(w,h) = \left\lceil \frac{A \cdot D_\mathrm{area}(w,h)}{C_\mathrm{net}} \right\rceil.
\end{equation}
The physical deployment is fixed at the dimensioning week $w_\mathrm{dim}$ and busy hour $h_\mathrm{peak}$ (e.g., $w_\mathrm{dim}=0$ to size for launch, or $w_\mathrm{dim}=W$ to over-provision for end-of-horizon demand). Let $G^\mathrm{RAN}(w_\mathrm{dim}, h_\mathrm{peak})$ denote the dimensioned cluster size. The CapEx is:
\begin{equation}
\label{eq:capex}
C = G^\mathrm{RAN}(w_\mathrm{dim}, h_\mathrm{peak}) \cdot \mathrm{Cost}_\mathrm{server}.
\quad [\$],
\end{equation}
The weekly profile $\lambda_\mathrm{RAN}(h)$, derived from empirical urban macro-cell measurements~\cite{barlacchi2015dataset}, is shown in Fig.~\ref{fig:daily_usage}.
Because $G^\mathrm{RAN}(w,h)$ already encodes both demand growth and hourly variation, we define the weekly \gls{opex} as:
\begin{equation}
\label{eq:opex_weekly}
O(w) = \sum_{h=0}^{23} G^\mathrm{RAN}(w,h) \cdot \frac{P_\mathrm{server} \cdot \mathrm{PUE}}{1000} \cdot c_\mathrm{elec},
\end{equation}
%
%and the total OpEx over the deployment lifetime is
%
% \begin{equation}
% \label{eq:opex}
% O = \sum_{w=0}^{W-1} O(w).
% \end{equation}
%
where $P_\mathrm{server}$ is the server TDP in watts and the factor $1/1000$ converts to kilowatts.

Because the cluster is dimensioned to a fixed peak, servers are partially idle at all other times and their spare capacity is available for \gls{ai} workloads.
The number of GPUs $G_\mathrm{free}(w,h)$ available for AI at week $w$, hour $h$ is:
\begin{equation}
\label{eq:g_free}
G_\mathrm{free}(w,h) = G_\mathrm{total} - G^\mathrm{RAN}(w,h)
\end{equation}
%
%where the $\max$ prevents negative values if late-horizon demand saturates the cluster; 
when $\rho_\mathrm{R}=1$ or $w=w_\mathrm{dim}$, this reduces to $G_\mathrm{free}(h) = G_\mathrm{total} - G^\mathrm{RAN}(w_\mathrm{dim}, h)$.

%
% The daily-average free fraction is then $\bar{f}_\mathrm{free} = 1 - \frac{1}{24}\sum_{t=0}^{23} \lambda_\mathrm{RAN}(t) = 1 - \bar{U}_\mathrm{RAN}$.
% From the empirical \gls{ran} profile, the primary peak occurs at 16:00 with a secondary maximum at 09:00, yielding $\bar{U}_\mathrm{RAN} = 0.498$ and a daily-average free fraction $\bar{f}_\mathrm{free} = 0.502$.

\section{LLM Demand Model}
\label{sec:llm_demand}

We model the \gls{llm} inference demand that can be served by the \gls{airan} cluster during off-peak \gls{ran} hours.
The workload is characterized by three parameters: the market token price $p_\mathrm{tok}$~[\$/token], the per-\gls{gpu} output throughput $T_\mathrm{GPU}$~[tok/s], and the maximum sustainable request concurrency per \gls{gpu} $\rho_\mathrm{max}$~[req/GPU].
All three depend on the specific \gls{llm} and hardware platform and are instantiated for a concrete model in Section~\ref{sec:scenarios}.
Following Xiao~et~al.~\cite{xiao2025densing}, we introduce a \emph{density scaling} parameter $\rho_\mathrm{dens}$ (annual capability density growth factor) that captures the growth of \gls{llm} capability density over time: the same hardware delivers more capable output as models improve.
The effective per-\gls{gpu} throughput at week $w$ is then
\begin{equation}
\label{eq:C_GPU_density}
T_\mathrm{GPU}(w) = T_\mathrm{GPU}(0) \cdot \rho_\mathrm{dens}^{w/52}
\quad [\text{tok/s}],
\end{equation}
where $T_\mathrm{GPU}(0)$ is the baseline throughput at deployment ($w=0$).

To characterize temporal demand patterns and per-request workload, we use the BurstGPT dataset~\cite{burstgpt2024}, a real-world trace of \gls{llm} inference requests containing timestamped records with input token counts.
From this trace we extract two quantities: (i)~a raw hourly request count $\lambda_\mathrm{LLM}(h)$ that captures the weekly shape of inference demand, and (ii)~the average number of tokens per request $\bar{T}_\mathrm{req}$.
Because the trace does not identify individual users, we cannot directly estimate per-user request rates from it.
Instead, we derive the baseline per-user daily request rate $\bar{q}(0)$ from public ChatGPT usage statistics: with approximately 2.5--3~billion prompts processed daily across 190.6~million daily active users~\cite{sqmagazine2025}, the ratio yields $\bar{q}(0) \approx 14.4$~req/user/day.

We model the per-user daily request rate as a time-varying quantity $\bar{q}(w)$ that compounds at the annual growth factor $\rho_\mathrm{LLM}$:
\begin{equation}
\label{eq:q_trend}
\bar{q}(w) = \bar{q}(0) \cdot \rho_\mathrm{LLM}^{w/52}
\quad [\text{req/user/day}],
\end{equation}
where $w$ is the week index.
We rescale the BurstGPT hourly profile so that its daily per-user average matches $\bar{q}(w)$.
Normalizing the raw counts gives the weekly shape $\hat{\lambda}_\mathrm{LLM}(h) = \lambda_\mathrm{LLM}(h) / \sum_{h=0}^{23} r(h)$, representing the fraction of daily requests that fall in hour~$h$.
The number of \gls{llm}-active users in the deployment area is $A \cdot \rho_\mathrm{pop} \cdot \eta_\mathrm{pen} \cdot \eta_\mathrm{AI}$.
The hourly request arrival rate at week $w$, hour $h$ is then
\begin{equation}
\label{eq:rps}
\Lambda(w,h) = \frac{\hat{\lambda}_\mathrm{LLM}(h)}{3600} \cdot \bar{q}(w) \cdot A \cdot \rho_\mathrm{pop} \cdot \eta_\mathrm{pen} \cdot \eta_\mathrm{AI}
\quad [\text{req/s}].
\end{equation}
The profile $\lambda_\mathrm{LLM}(h)$ is shown alongside $\lambda_\mathrm{RAN}(h)$ in Fig.~\ref{fig:daily_usage}.
The two curves are \emph{partially anti-correlated}: the morning \gls{ran} peak coincides with the \gls{llm} demand trough, and the evening \gls{llm} peak coincides with declining \gls{ran} load, with the primary conflict window between 14:00 and 18:00.
This complementarity is favorable for dual-use server scheduling (without considering de-synchronized curves based on different markets or geographical areas).

Given that each request requires on average $\bar{T}_\mathrm{req}$ tokens to process, the mean service time per request at week $w$ is $\bar{s}(w) = \bar{T}_\mathrm{req} / T_\mathrm{GPU}(w)$.
Applying Little's Law to the inference queue, the mean number of concurrently active requests at week $w$, hour $h$ is $L(w,h) = \Lambda(w,h) \cdot \bar{s}(w)$, and the number of \glspl{gpu} required to sustain the full demand is:
\begin{equation}
\label{eq:g_LLM_req}
G^\mathrm{LLM}_\mathrm{req}(w,h) = \left\lceil \frac{L(w,h)}{\rho_\mathrm{max}} \right\rceil
= \left\lceil \frac{\Lambda(w,h) \cdot \bar{T}_\mathrm{req}}{T_\mathrm{GPU}(w) \cdot \rho_\mathrm{max}} \right\rceil
\end{equation}
Specific parameter values and the user scenarios evaluated in this paper are detailed in Section~\ref{sec:scenarios}.

\section{Revenue Model for AI-on-RAN}
\label{sec:roi}

The \gls{airan} revenue model combines the \gls{ran} deployment cost (Section~\ref{sec:ran_demand}) with the incremental income generated by renting surplus compute to \gls{llm} inference consumers.
At each week $w$, hour $h$, the number of GPUs available for \gls{ai} workloads is $G_\mathrm{free}(w,h)$ (Eq.~\eqref{eq:g_free}).
The LLM allocation $G^\mathrm{LLM}_\mathrm{alloc}(w,h)$ is capped by the instantaneous \gls{llm} demand from Section~\ref{sec:llm_demand}:
\begin{equation}
\label{eq:g_LLM_alloc}
G^\mathrm{LLM}_\mathrm{alloc}(w,h) = \min\!\bigl(G^\mathrm{LLM}_\mathrm{req}(w,h),\; G_\mathrm{free}(w,h)\bigr).
\end{equation}
The resulting token throughput delivered to inference consumers at week $w$, hour $h$ is:
\begin{equation}
\label{eq:tok_throughput}
T(w,h) = G^\mathrm{LLM}_\mathrm{alloc}(w,h) \cdot \rho_\mathrm{max} \cdot T_\mathrm{GPU}(w)
\quad [\text{tok/s}].
\end{equation}
Tokens are priced at a market rate $p_\mathrm{tok}$ [\$/token].
Motivated by the empirical price trends documented by Demirer et al.~\cite{demirer2025market}, we model token price erosion through a depreciation factor $\rho_\mathrm{tok} \in (0,1]$, where : $p_\mathrm{tok}(w) = p_\mathrm{tok} \cdot \rho_\mathrm{tok}^{w/52}$;
Over a deployment lifetime of $W$ weeks, the total \gls{llm} revenue is thus:
\begin{equation}
\label{eq:llm_revenue_weekly}
R_\mathrm{LLM}(w) = 7 \cdot p_\mathrm{tok}(w) \cdot \sum_{h=0}^{23} T(w,h) \cdot 3600
\end{equation}
and the total \gls{llm} revenue over the deployment lifetime is
\begin{equation}
\label{eq:llm_revenue}
R_\mathrm{LLM} = \sum_{w=0}^{W-1} R_\mathrm{LLM}(w).
\end{equation}
The net economic gain of the \gls{airan} investment over a conventional \gls{ran}-only deployment is then the difference between $R_\mathrm{LLM}$ and the additional CapEx premium of \gls{gpu} platforms over dedicated \gls{fec}-accelerator alternatives, amortized over the deployment lifetime.
Scenario-specific parameter choices and the resulting revenue projections are presented in Section~\ref{sec:scenarios}.

\section{Scenario-Based Evaluation}
\label{sec:scenarios}

\begin{table}[t]
    \centering
    \caption{Scenario parameters -- Milan dense urban deployment.}
    \label{tab:milan_params}
    \begin{tabular}{llrl}
        \toprule
        \textbf{Parameter} & \textbf{Symbol} & \textbf{Value} & \textbf{Ref.} \\
        \midrule
        \multicolumn{4}{c}{\textit{\gls{ran} demand}} \\
        Population density & $\rho_\mathrm{pop}$ & 7500~/km$^2$ & \cite{istat2024} \\
        Smartphone penetration & $\eta_\mathrm{pen}$ & 80\% & \cite{gsma2024} \\
        Busy-hour concurrency & $\alpha_\mathrm{BH}$ & 10\% & \cite{smartcities7060140} \\
        Per-user downlink rate  & $R_\mathrm{user}(0)$ & 300~Mbps & \cite{itu2160} \\
        Avg.\ spectral efficiency & $\mathrm{SE}$ & 9~bit/s/Hz & \cite{itu2030wd1} \\
        L1/L2 overhead & $\eta_\mathrm{OH}$ & 20\% & \cite{3gpp38214} \\
        Power Usage Effectiveness  & $\mathrm{PUE}$ & 1.5 & \cite{uptime2024} \\
        Electricity cost & $c_\mathrm{elec}$ & 0.04~\$/kWh & \cite{eurostat2024elec} \\
        RAN demand growth factor & $\rho_\mathrm{R}$ & 1.2 & \cite{ericsson2025mobility} \\
        \midrule
        % \vspace{0.1em}\\
        \multicolumn{4}{c}{\textit{\gls{llm} inference (Llama~3.1 70B \gls{fp8})}} \\
        Per-\gls{gpu} output throughput & $T_\mathrm{GPU}(0)$ & 37~tok/s & \cite{artificialanalysis2024} \\
        %Density scaling  & $\rho_\mathrm{dens}$ & --- & \cite{xiao2025densing} \\
        Max concurrency per \gls{gpu} & $\rho_\mathrm{max}$ & 23.5~req/GPU & \cite{pichlmeier2024} \\
        Avg.\ tokens per request & $\bar{T}_\mathrm{req}$ & 969.17 & \cite{burstgpt2024} \\
        Per-user daily requests  & $\bar{q}(0)$ & 14.4~req/day & \cite{sqmagazine2025} \\
        Token price & $p_\mathrm{tok}$ & 0.88~\$/Mtok & \cite{together2024} \\
        %Token depreciation  & $\rho_\mathrm{tok}$ & 0.5 & \cite{demirer2025market} \\
        \gls{llm} adoption ratio & $\eta_\mathrm{AI}$ & 50\% & \cite{savoldi2025generative} \\
        LLM demand growth factor & $\rho_\mathrm{LLM}$ & 16 & \cite{demirer2025market} \\
        \bottomrule
    \end{tabular}
\end{table}

We instantiate the models of Sections~\ref{sec:ran_demand}--\ref{sec:roi} for a dense urban deployment in Milan, Italy.
Table~\ref{tab:milan_params} summarizes the \gls{ran} and \gls{llm} parameters used across all scenarios.
Throughout the evaluation, we consider two dimensioning strategies for the \gls{airan} cluster:
\begin{itemize}
    \item \textbf{Scenario~1}: the cluster is sized for launch demand, $w_\mathrm{dim}=0$, with \gls{ran} demand held constant over the 10-year lifetime ($\rho_\mathrm{R}=1$).
    \item\textbf{Scenario~2}: the cluster is sized for end-of-horizon demand, $w_\mathrm{dim}=W$, with \gls{ran} demand growing at $\rho_\mathrm{R}=1.2$.
\end{itemize}

\begin{figure}
  \centering
    \setlength{\abovecaptionskip}{-.1cm}
  \input{figures/gpu_usage_pgfplot}
  \caption{Hourly \gls{gpu} allocation at deployment ($w=0$) for clusters sized for Scenario~1 (top) and Scenario~2 (bottom).
  The total deployed capacity $G_\mathrm{total}$ is split at each hour between \gls{ran} processing and \gls{llm} inference.}
  \label{fig:gpu_dim0}\label{fig:gpu_dim10}
\end{figure}
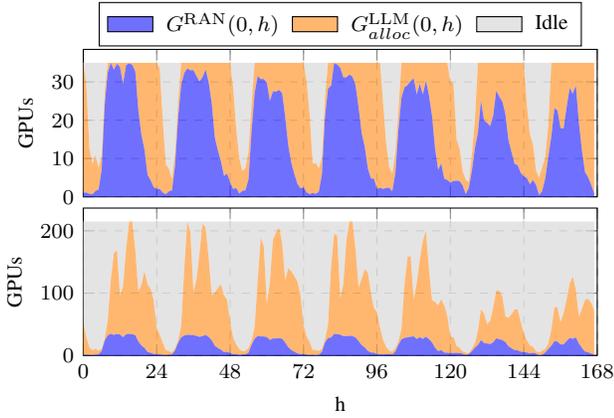

\subsection{GPUs Allocation}

Figure~\ref{fig:gpu_dim0} illustrates the hourly \gls{gpu} allocation at deployment ($w=0$) under both dimensioning strategies, when considering the NVIDIA Aerial deployment.
In Scenario~1 (top panel), the cluster is sized to cover the weekly busy-hour peak, so little surplus remains for the \gls{llm} tenant; inference revenue is thus limited to hours when the \gls{ran} load is low.
In Scenario~2 (bottom panel), the cluster is provisioned for end-of-horizon demand.
At $w=0$, \gls{ran} traffic is well below the dimensioning point, so the workload occupies only a fraction of the deployed capacity at any hour, and the surplus $G_\mathrm{free}(0,h)$ is large---particularly during off-peak hours.
As $w$ increases toward $w_\mathrm{dim}$, \gls{ran} demand absorbs a growing share of this surplus, progressively reducing the headroom available for \gls{ai} tenants.

\begin{figure}
  \centering
    \setlength{\abovecaptionskip}{-.1cm}
  \input{figures/gpu_allocation_pgfplot}
  \caption{Weekly-averaged \gls{gpu} allocation (\gls{ran} plus \gls{llm}) over the 10-year horizon.
  Top: Scenario~1.  Bottom: Scenario~2 .}
  \label{fig:gpu_allocation_trend}
\end{figure}
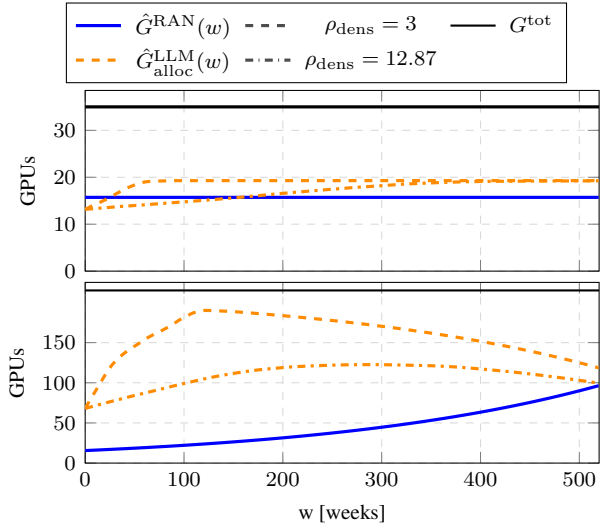

In order to evaluate the weekly-averaged GPU allocation, we define $\hat{G}(w)=1/168\cdot\sum_h{G}(w,h)$, where 168 is the number of hours in a week. $\hat{G}(w)$ is traced for both \gls{llm} and \gls{ran} in Figure~\ref{fig:gpu_allocation_trend} across the full 10-year horizon.
In Scenario~1, \gls{ran} demand is constant and consumes on average 15~\glspl{gpu} throughout.
In Scenario~2, the dynamics are richer. At deployment, the \gls{ran} workload occupies about 15~\glspl{gpu}; as traffic grows at 20\%/year, its allocation rises to roughly 100~\glspl{gpu} by the end of the horizon.

For the \gls{llm} metric, different values of $\rho_{dens}$ are explored, as there is still no consensus of what this ratio is and will be. The first one ($\rho_{dens}=3$) has been estimated by Gundlach et Al, based on the price reduction\cite{gundlach2025price}, while the second one ($\rho_{dens}=12.87$) has been estimated by \cite{xiao2025densing} by fitting a exponential model on the performance of different \glspl{llm}.
% The \gls{llm} workload starts at around 145~\glspl{gpu}, peaks near 155 after four years, and then declines as the \gls{ran} workload reclaims capacity.
The $\rho_\mathrm{dens}$ parameter significantly modulates both the peak allocation and its timing.
Under $\rho_\mathrm{dens}=3$, slower performance densification translates into a steeper demand curve: in Scenario~2, the \gls{llm} allocation reaches approximately 190~\glspl{gpu} as early as year two and a half, before declining as the \gls{ran} workload progressively absorbs the shared capacity.
With $\rho_\mathrm{dens}=12.87$, efficiency gains more effectively offset demand growth, keeping the peak near 122~\glspl{gpu} around year five and producing a gentler subsequent decline.
In Scenario~1, the contrast manifests in the rate of saturation rather than the ceiling: the low-densification case fills the available excess capacity in roughly two years, while the high-densification case approaches the same ceiling only gradually over the full ten-year horizon.

\subsection{LLM Revenues}
To characterize the interplay between token price deflation and performance densification, we define the ratio $k = \rho_\mathrm{tok}/\rho_\mathrm{dens}$, which captures the relative speed at which token prices erode with respect to the efficiency gains achieved through model densification.
In practice, these two quantities are structurally coupled: the market-observed token price deflation is largely driven by the very efficiency improvements that $\rho_\mathrm{dens}$ tracks, making their ratio the natural quantity to parametrize.
We evaluated the revenue metric under both $\rho_\mathrm{dens}$ values considered above ($\rho_\mathrm{dens}=3$ and $\rho_\mathrm{dens}=12.87$) and found only marginal differences in the resulting revenue trajectories; accordingly, the remainder of the analysis fixes $\rho_\mathrm{dens}=12.87$ and sweeps $k$ directly.

\begin{figure}
  \centering
  \input{figures/llm_revenue_pgfplot}
  \setlength{\abovecaptionskip}{-.1cm}
  \caption{Weekly \gls{llm} gross revenue over the deployment lifetime under different values of $k=\rho_\mathrm{tok}/\rho_\mathrm{dens}$ (token depreciation relative to efficiency improvement).
  Top: Scenario~1.
  Bottom: Scenario~2.}
  \label{fig:llm_revenue}
\end{figure}
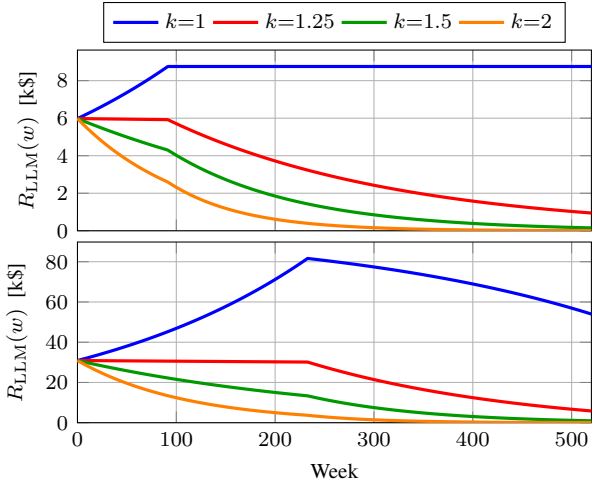

Figure~\ref{fig:llm_revenue} illustrates how \gls{llm} revenue evolves over the deployment horizon across different values of $k$.
The case $k{=}1$ represents the neutral point at which token price deflation is exactly offset by efficiency gains, so the effective per-token price remains constant over time.
For $k{>}1$, deflation outpaces efficiency improvement and the effective price erodes monotonically.
In Scenario~1 (top panel), all curves start at approximately \$6000/week.
The $k{=}1$ case grows steadily as the \gls{llm} workload fills the available excess capacity, reaching a plateau of roughly \$8750/week after about 92~weeks; beyond this inflection, the curve remains flat for the rest of the horizon since neither the effective token price nor the \gls{gpu} allocation change further.
For $k{>}1$, however, revenue declines from the very first week: token price deflation erodes earnings faster than the growing \gls{gpu} allocation can compensate.
The decline accelerates once the capacity headroom is exhausted at the same inflection point, since the partial offsetting effect of additional \gls{gpu} hours disappears.
By the end of the ten-year horizon, the $k{=}2$ curve is reduced to a negligible fraction of its initial value, with $k{=}1.5$ and $k{=}1.25$ following the same qualitative trend at progressively slower rates.

In Scenario~2 (bottom panel), starting from approximately \$31000/week, the $k{=}1$ curve grows continuously for roughly five years---peaking near \$80000/week around week~257---before declining as the expanding \gls{ran} workload progressively reclaims shared \gls{gpu} hours.
For $k{>}1$, revenue falls from the outset; beyond week~257 the loss of \gls{gpu} capacity further compounds the token price erosion, steepening the decline for all curves.
Crucially, the $k{=}2$ curve approaches zero by the end of the horizon, revealing a structural threshold: when token price deflation outpaces efficiency gains by a factor of two, \gls{llm} operation becomes economically non-viable regardless of the available infrastructure.

\subsection{ROI Analysis}

To determine whether the \gls{llm} co-tenant justifies the \gls{airan} cost premium, we frame the problem as a marginal investment analysis.
Let $C^\mathrm{Aerial}$ and $C^\mathrm{FlexRAN}$ denote the CapEx of each platform (Eq.~\eqref{eq:capex}).
The marginal investment is the additional CapEx required to deploy Aerial over the \gls{fec}-accelerated alternative:
\begin{equation}
\label{eq:investment}
I = C^\mathrm{Aerial} - C^\mathrm{FlexRAN}.
\end{equation}
The cumulative net return is the gross \gls{llm} revenue minus the Aerial operational cost over the deployment horizon:
\begin{equation}
\label{eq:return}
R(w) = R_\mathrm{LLM}(w) - O(w)^\mathrm{Aerial}.
\end{equation}
Break-even is reached when cumulative $R$ equals $I$; the ratio $R/I$ quantifies the return multiple over the full horizon.

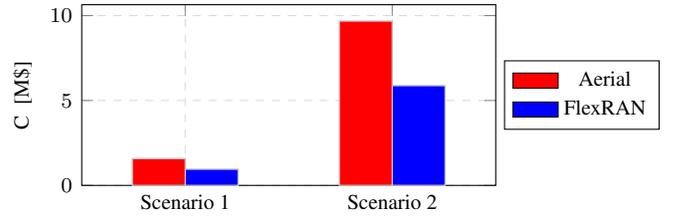
\begin{figure}
  \centering
  \input{figures/milan_tco_pgfplot}
    \setlength{\abovecaptionskip}{-.1cm}
  \caption{CapEx for the Milan network for Aerial and FlexRAN under Scenario~1 and Scenario~2.}
  \label{fig:textcomp}
\end{figure}

Figure~\ref{fig:textcomp} shows the CapEx breakdown for Aerial and FlexRAN under both dimensioning strategies.
We focus on CapEx here because the OpEx depends on the actual \gls{llm} demand: under \gls{ran}-only operation, servers are power-gated during low-traffic periods (e.g., overnight), so the incremental energy expenditure is directly tied to the \gls{llm} workload and is therefore accounted for in the net return $R$ (Eq.~\eqref{eq:return}).
In Scenario~1, Aerial requires a CapEx of \$1.58M against \$0.95M for FlexRAN, placing the marginal investment at $I \approx \$0.62$M.
In Scenario~2, the gap widens substantially: \$9.68M versus \$5.87M, yielding $I \approx \$3.80$M.
This premium is a direct consequence of the \gls{gpu}-based server architecture, whose per-unit acquisition cost significantly exceeds that of the \gls{fpga}-accelerated FlexRAN hardware.
Transitioning from Scenario~1 to Scenario~2 increases CapEx by 514\% for Aerial and 518\% for FlexRAN, since the cluster must be sized for end-of-horizon demand from day one.
The trade-off is that this over-provisioning creates a large pool of idle \gls{gpu} hours from the outset, substantially enlarging the window for \gls{llm} revenue generation throughout the deployment lifetime.

\begin{figure}
  \centering
  \input{figures/llm_revenue_cumulative_pgfplot}
    \setlength{\abovecaptionskip}{-.1cm}
  \caption{Cumulative \gls{llm} revenue $R$ (Eq.~\eqref{eq:return}) over the deployment horizon under different values of $k = \rho_\mathrm{tok}/\rho_\mathrm{dens}$.
  Top: Scenario~1.
  Bottom: Scenario~2.
  The horizontal dashed line indicates the marginal investment $I$ (Eq.~\eqref{eq:investment}).}
  \label{fig:llm_revenue_cumulative}
\end{figure}
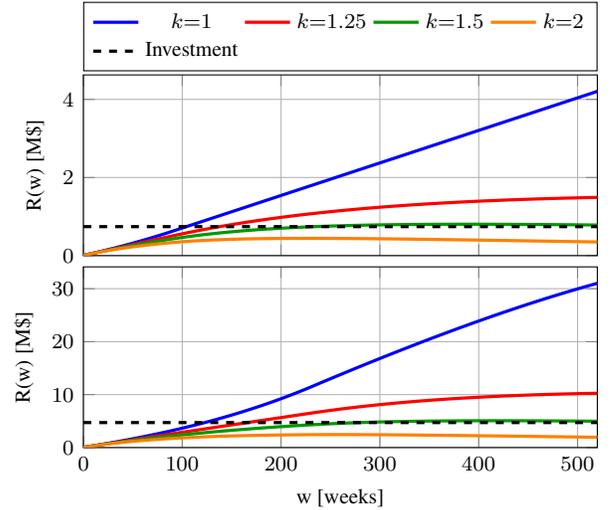

Figure~\ref{fig:llm_revenue_cumulative} plots cumulative revenue alongside the investment threshold for each scenario.
The qualitative pattern mirrors the weekly revenue analysis: curves with lower $k$ accumulate faster and recover the investment sooner, while higher $k$ leads to slower accumulation or outright failure to break even.
In Scenario~1 (top panel), the $k{=}1$ curve crosses the threshold at approximately week~105 (two years into deployment), followed by $k{=}1.25$ around week~139.
The $k{=}1.5$ curve requires nearly the full horizon to recover the investment, crossing only around week~235.
The $k{=}2$ curve never reaches break-even: the cumulative revenue at week~520 (\$0.35M) falls far short of the \$0.62M threshold, confirming that token deflation at twice the rate of efficiency improvement makes the investment irrecoverable on this timescale.
In Scenario~2 (bottom panel), the investment threshold is substantially higher at \$3.80M, yet the larger volume of idle \gls{gpu} capacity available from the start allows all curves except $k{=}2$ to cross it within the horizon.
The $k{=}1$ curve breaks even around week~123, $k{=}1.25$ around week~166, and $k{=}1.5$ recovers the investment near week~280.
As in Scenario~1, the $k{=}2$ case fails to break even: despite generating \$1.94M in cumulative revenue, it does not reach the \$3.80M threshold, as the compounding effect of token deflation and shrinking \gls{gpu} availability forecloses recovery.

\begin{figure}[H]
  \centering
  \input{figures/roi_pgfplot}
    \setlength{\abovecaptionskip}{-.3cm}
  \caption{Return on investment ($R/I$) of \gls{airan} by scenario and depreciation ratio $k = \rho_\mathrm{tok}/\rho_\mathrm{dens}$.
  Investment $I$ and return $R$ defined in Eqs.~\eqref{eq:investment} and \eqref{eq:return}.}
  \label{fig:roi}
\end{figure}
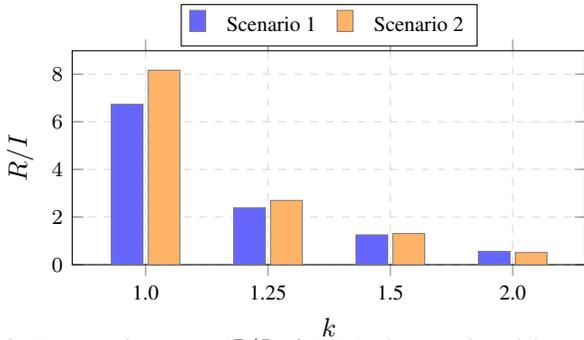

Figure~\ref{fig:roi} summarises the final return multiple $R/I$ across all depreciation scenarios.
The central finding is that Scenario~2 consistently outperforms Scenario~1 in all cases where the investment is profitable: at $k{=}1$, the return multiple reaches $8.2{\times}$ in Scenario~2 versus $6.7{\times}$ in Scenario~1, and at $k{=}1.25$ the figures are $2.7{\times}$ and $2.4{\times}$, respectively.
This advantage stems directly from the deliberate over-provisioning of Scenario~2: by dimensioning the cluster for end-of-horizon demand, the operator creates an infrastructure surplus that can be leased for \gls{llm} inference from day one, generating revenue that far exceeds the additional CapEx.
At $k{=}1.5$, both scenarios remain marginally profitable ($R/I \approx 1.31{\times}$ and $1.25{\times}$), confirming that moderate token deflation can still be offset by the efficiency gains embedded in $k$.
At $k{=}2$, both scenarios yield $R/I < 1$ (0.56${\times}$ and 0.51${\times}$), establishing a hard threshold: when token price erosion outpaces efficiency improvement by a factor of two, the \gls{llm} co-tenancy fails to recover the infrastructure premium regardless of the dimensioning strategy.
Notably, the $k{=}2$ return is slightly lower in Scenario~2 than in Scenario~1, because the larger investment denominator more than offsets the additional revenue generated by the excess capacity.

\section{Conclusion}
\label{sec:conclusion}
This paper presented a techno-economic framework for evaluating AI-RAN deployments in which \gls{gpu}-accelerated baseband platforms share idle capacity with \gls{llm} inference workloads.
By combining publicly available hardware benchmarks, IMT-2030-anchored demand models, and empirical \gls{llm} traffic traces, we constructed a unified cost and revenue model and instantiated it for a dense urban deployment in Milan, Italy.
The analysis compared two dimensioning strategies:sizing the cluster for launch demand (Scenario~1) and for end-of-horizon demand (Scenario~2); across a range of token depreciation-to-densification ratios~$k$.

Three principal findings emerge.
For all $k \leq 1.5$, the cumulative inference revenue exceeds the marginal cost premium of the \gls{gpu}-based platform, with return multiples up to $8.2\times$.
Deliberate over-provisioning (Scenario~2) consistently outperforms conservative dimensioning, as the early-year surplus generates substantial revenue before RAN demand absorbs the excess capacity.
The ratio $k = \rho_\text{tok}/\rho_\text{dens}$ emerges as the decisive parameter: at $k \geq 2$, the investment becomes irrecoverable regardless of dimensioning strategy, establishing a clear viability boundary for AI-RAN economics.

Several limitations of the current analysis suggest directions for future work.
The revenue model prices inference tokens at prevailing cloud market rates and does not capture the additional value that edge-local inference may command: lower latency for interactive applications, reduced backhaul and transport costs for both the RAN operator and the \gls{llm} tenant, and data locality advantages for privacy-sensitive workloads.
Incorporating an edge premium into the token price would likely strengthen the economic case further.
The \gls{llm} demand model assumes a single workload type (\glspl{llm}); incorporating heterogeneous AI workloads such as image generation, embedding computation, or fine-tuning jobs would broaden the revenue base and potentially smooth the demand profile.
Finally, the analysis also assumes a single geographic market with synchronized RAN and \gls{llm} diurnal patterns; in practice, operators serving multiple time zones or offering compute to geographically dispersed AI tenants could exploit additional temporal diversity.
Despite these simplifications, the framework provides a principled and extensible basis for operators and policymakers to assess the economic viability of accelerator-based RAN architectures in the transition to~6G.

\bibliographystyle{IEEEtran}

\bibliography{AI_RAN_economy}

\end{document}

%% file: acronyms.tex
\newacronym{aoa}{AoA}{Angle of Arrival}
\newacronym{dapp}{dApp}{distributed Application}
\newacronym{cuphy}{cuPHY}{CUDA Physical layer}
\newacronym{cuda}{CUDA}{Compute Unified Device Architecture}
\newacronym{cubb}{cuBB}{CUDA Baseband}
\newacronym{onnx}{ONNX}{Open Neural Network Exchange}
\newacronym{ldpc}{LDPC}{Low Density Parity-Check Code}
\newacronym{crc}{CRC}{Cyclic Redundancy Check}
\newacronym{dnn}{DNN}{Deep Neural Network}
\newacronym{ort}{ORT}{ONNX Runtime}
\newacronym{tensorrt}{TensorRT}{Tensor RealTime}
\newacronym{cnn}{CNN}{Convolutional Neural Network}
\newacronym{cudnn}{cuDNN}{CUDA Deep Neural Network library}
\newacronym{cublas}{cuBLAS}{CUDA Basic Linear Algebra Subroutines}
\newacronym{hdf5}{HDF5}{Hierarchical Data Format version 5}
\newacronym{rmr}{RMR}{RIC Message Router}
\newacronym{3gpp}{3GPP}{3rd Generation Partnership Project}
\newacronym{4g}{4G}{4th generation}
\newacronym{5g}{5G}{fifth generation}
\newacronym{6g}{6G}{sixth generation}
\newacronym{5gc}{5GC}{5G Core}
\newacronym{adc}{ADC}{Analog to Digital Converter}
\newacronym{aerpaw}{AERPAW}{Aerial Experimentation and Research Platform for Advanced Wireless}
\newacronym{ai}{AI}{Artificial Intelligence}
\newacronym{aimd}{AIMD}{Additive Increase Multiplicative Decrease}
\newacronym{am}{AM}{Acknowledged Mode}
\newacronym{amc}{AMC}{Adaptive Modulation and Coding}
\newacronym{amf}{AMF}{Access and Mobility Management Function}
\newacronym{aops}{AOPS}{Adaptive Order Prediction Scheduling}
\newacronym{api}{API}{Application Programming Interface}
\newacronym{apn}{APN}{Access Point Name}
\newacronym{aqm}{AQM}{Active Queue Management}
\newacronym{arc-ota}{ARC-OTA}{Aerial RAN CoLab Over-the-Air}
\newacronym{adl}{ADL}{Aerial Data Lake}
\newacronym{ausf}{AUSF}{Authentication Server Function}
\newacronym{avc}{AVC}{Advanced Video Coding}
\newacronym{awgn}{AWGN}{Additive White Gaussian Noise}
\newacronym{balia}{BALIA}{Balanced Link Adaptation Algorithm}
\newacronym{bbu}{BBU}{Base Band Unit}
\newacronym{bdp}{BDP}{Bandwidth-Delay Product}
\newacronym{ber}{BER}{Bit Error Rate}
\newacronym{bf}{BF}{Beamforming}
\newacronym{bler}{BLER}{Block Error Rate}
\newacronym{brr}{BRR}{Bayesian Ridge Regressor}
\newacronym{bsr}{BSR}{Buffer Status Report}
\newacronym{bs}{BS}{Base Station}
\newacronym{bpsk}{BPSK}{Binary Phase-shift keying}
\newacronym{bss}{BSS}{Business Support System}
\newacronym{ca}{CA}{Carrier Aggregation}
\newacronym{caas}{CaaS}{Connectivity-as-a-Service}
\newacronym{cb}{CB}{Code Block}
\newacronym{cc}{CC}{Congestion Control}
\newacronym{ccid}{CCID}{Congestion Control ID}
\newacronym{cco}{CC}{Carrier Component}
\newacronym{cdd}{CDD}{Cyclic Delay Diversity}
\newacronym{cdf}{CDF}{Cumulative Distribution Function}
\newacronym{cdn}{CDN}{Content Distribution Network}
\newacronym{cir}{CIR}{Channel Impulse Response}
\newacronym{cn}{CN}{Core Network}
\newacronym{codel}{CoDel}{Controlled Delay Management}
\newacronym{comac}{COMAC}{Converged Multi-Access and Core}
\newacronym{cord}{CORD}{Central Office Re-architected as a Datacenter}
\newacronym{cornet}{CORNET}{COgnitive Radio NETwork}
\newacronym{cosmos}{COSMOS}{Cloud Enhanced Open Software Defined Mobile Wireless Testbed for City-Scale Deployment}
\newacronym{cots}{COTS}{Commercial Off-the-Shelf}
\newacronym{cp}{CP}{Control Plane}
\newacronym{cpu}{CPU}{Central Processing Unit}
\newacronym{cqi}{CQI}{Channel Quality Information}
\newacronym{cr}{CR}{Cognitive Radio}
\newacronym{cran}{CRAN}{Cloud \gls{ran}}
\newacronym{crs}{CRS}{Cell Reference Signal}
\newacronym{csi}{CSI}{Channel State Information}
\newacronym{csirs}{CSI-RS}{Channel State Information - Reference Signal}
\newacronym{cu}{CU}{Central Unit}
\newacronym{d2tcp}{D$^2$TCP}{Deadline-aware Data center TCP}
\newacronym{d3}{D$^3$}{Deadline-Driven Delivery}
\newacronym{dac}{DAC}{Digital to Analog Converter}
\newacronym{dag}{DAG}{Directed Acyclic Graph}
\newacronym{darpa}{DARPA}{Defense Advanced Research Projects Agency}
\newacronym{das}{DAS}{Distributed Antenna System}
\newacronym{dash}{DASH}{Dynamic Adaptive Streaming over HTTP}
\newacronym{dc}{DC}{Direct Current}
\newacronym{dccp}{DCCP}{Datagram Congestion Control Protocol}
\newacronym{dce}{DCE}{Direct Code Execution}
\newacronym{dci}{DCI}{Downlink Control Information}
\newacronym{dcl}{DCL}{Dear Colleague Letter}
\newacronym{dctcp}{DCTCP}{Data Center TCP}
\newacronym{dl}{DL}{Downlink}
\newacronym{dmr}{DMR}{Deadline Miss Ratio}
\newacronym{dmrs}{DMRS}{DeModulation Reference Signal}
\newacronym{dpu}{DPU}{Data Processing Unit}
\newacronym{drlcc}{DRL-CC}{Deep Reinforcement Learning Congestion Control}
\newacronym{drs}{DRS}{Discovery Reference Signal}
\newacronym{du}{DU}{Distributed Unit}
\newacronym{e2e}{E2E}{end-to-end}
\newacronym{e2sm}{E2SM}{E2 Service Model}
\newacronym{e3ap}{E3AP}{E3 Application Protocol}
\newacronym{e3sm}{E3SM}{E3 Service Model}
\newacronym{ecaas}{ECaaS}{Edge-Cloud-as-a-Service}
\newacronym{ecn}{ECN}{Explicit Congestion Notification}
\newacronym{edf}{EDF}{Earliest Deadline First}
\newacronym{eirp}{EIRP}{Effective Isotropic Radiated Power}
\newacronym{em}{EM}{Electro-Magnetic}
\newacronym{embb}{eMBB}{Enhanced Mobile Broadband}
\newacronym{empower}{EMPOWER}{EMpowering transatlantic PlatfOrms for advanced WirEless Research}
\newacronym{enb}{eNB}{evolved Node Base}
\newacronym{endc}{EN-DC}{E-UTRAN-\gls{nr} \gls{dc}}
\newacronym{epc}{EPC}{Evolved Packet Core}
\newacronym{eps}{EPS}{Evolved Packet System}
\newacronym{es}{ES}{Edge Server}
\newacronym{etsi}{ETSI}{European Telecommunications Standards Institute}
\newacronym[firstplural=Estimated Times of Arrival (ETAs)]{eta}{ETA}{Estimated Time of Arrival}
\newacronym{eutran}{E-UTRAN}{Evolved Universal Terrestrial Access Network}
\newacronym{faas}{FaaS}{Function-as-a-Service}
\newacronym{fapi}{FAPI}{Functional Application Platform Interface}
\newacronym{fcc}{FCC}{Federal Communications Commission}
\newacronym{fdd}{FDD}{Frequency Division Duplexing}
\newacronym{fdm}{FDM}{Frequency Division Multiplexing}
\newacronym{fdma}{FDMA}{Frequency Division Multiple Access}
\newacronym{fed4fire}{FED4FIRE+}{Federation 4 Future Internet Research and Experimentation Plus}
\newacronym{fir}{FIR}{Finite Impulse Response}
\newacronym{fit}{FIT}{Future \acrlong{iot}}
\newacronym{fpga}{FPGA}{Field Programmable Gate Array}
\newacronym{fps}{FPS}{Frames per second}
\newacronym{fr1}{FR1}{Frequency Range 1}
\newacronym{fr2}{FR2}{Frequency Range 2}
\newacronym{fs}{FS}{Fast Switching}
\newacronym{fscc}{FSCC}{Flow Sharing Congestion Control}
\newacronym{ftp}{FTP}{File Transfer Protocol}
\newacronym{fw}{FW}{Flow Window}
\newacronym{ga128}{Ga}{Golay Sequence type A}
\newacronym{ge}{GE}{Gaussian Elimination}
\newacronym{gh}{GH}{Grace Hopper}
\newacronym{glfsr}{GLFSR}{Galois Linear Feedback Shift Register}
\newacronym{gnb}{gNB}{Next Generation Node B}
\newacronym{gold}{Gold}{Gold}
\newacronym{gop}{GOP}{Group of Pictures}
\newacronym{gpr}{GPR}{Gaussian Process Regressor}
\newacronym{gpu}{GPU}{Graphics Processing Unit}
\newacronym{grpc}{gRPC}{gRPC Remote Procedure Calls}
\newacronym{gtp}{GTP}{GPRS Tunneling Protocol}
\newacronym{gtpc}{GTP-C}{GPRS Tunnelling Protocol Control Plane}
\newacronym{gtpu}{GTP-U}{GPRS Tunnelling Protocol User Plane}
\newacronym{gtpv2c}{GTPv2-C}{\gls{gtp} v2 - Control}
\newacronym{gw}{GW}{Gateway}
\newacronym{harq}{HARQ}{Hybrid Automatic Repeat Request}
\newacronym{hest}{$\hat{\mathbf{H}}$}{Channel Estimates}
\newacronym{hetnet}{HetNet}{Heterogeneous Network}
\newacronym{hh}{HH}{Hard Handover}
\newacronym{hol}{HOL}{Head-of-Line}
\newacronym{hqf}{HQF}{Highest-quality-first}
\newacronym{hss}{HSS}{Home Subscription Server}
\newacronym{http}{HTTP}{HyperText Transfer Protocol}
\newacronym{ia}{IA}{Initial Access}
\newacronym{iab}{IAB}{Integrated Access and Backhaul}
\newacronym{ic}{IC}{Incident Command}
\newacronym{ietf}{IETF}{Internet Engineering Task Force}
\newacronym{ifw}{IFW}{Interference Free Window}
\newacronym{imsi}{IMSI}{International Mobile Subscriber Identity}
\newacronym{imt}{IMT}{International Mobile Telecommunication}
\newacronym{iot}{IoT}{Internet of Things}
\newacronym{ip}{IP}{Internet Protocol}
\newacronym{iq}{I/Q}{In-phase and Quadrature}
\newacronym{isac}{ISAC}{Integrated Sensing and Communication}
\newacronym{itu}{ITU}{International Telecommunication Union}
\newacronym{iw}{IW}{Interference Whitening}
\newacronym{kl}{KL}{Kullback–Leibler}
\newacronym{kpi}{KPI}{Key Performance Indicator}
\newacronym{kpm}{KPM}{Key Performance Measurement}
\newacronym{kvm}{KVM}{Kernel-based Virtual Machine}
\newacronym{leo}{LEO}{Low Earth Orbit}
\newacronym{los}{LOS}{Line-of-Sight}
\newacronym{ls}{LS}{Loosely Synchronised}
\newacronym{lsm}{LSM}{Link-to-System Mapping}
\newacronym{lstm}{LSTM}{Long Short Term Memory}
\newacronym{lte}{LTE}{Long Term Evolution}
\newacronym{lxc}{LXC}{Linux Container}
\newacronym{m2m}{M2M}{Machine to Machine}
\newacronym{mac}{MAC}{Medium Access Control}
\newacronym{manet}{MANET}{Mobile Ad Hoc Network}
\newacronym{mano}{MANO}{Management and Orchestration}
\newacronym{mc}{MC}{Multi-Connectivity}
\newacronym{mcc}{MCC}{Mobile Cloud Computing}
\newacronym{mchem}{MCHEM}{Massive Channel Emulator}
\newacronym{mcs}{MCS}{Modulation and Coding Scheme}
\newacronym{mec}{MEC}{Multi-access Edge Computing}
\newacronym{mec2}{MEC}{Mobile Edge Cloud}
\newacronym{mfc}{MFC}{Mobile Fog Computing}
\newacronym{mi}{MI}{Mutual Information}
\newacronym{mig}{MIG}{Multi-Instance GPU}
\newacronym{mib}{MIB}{Master Information Block}
\newacronym{miesm}{MIESM}{Mutual Information Based Effective SINR}
\newacronym{mimo}{MIMO}{Multiple Input, Multiple Output}
\newacronym{mgen}{MGEN}{Multi-Generator}
\newacronym{ml}{ML}{Machine Learning}
\newacronym{mlp}{MLP}{Multilayer Perceptron}
\newacronym{mlr}{MLR}{Maximum-local-rate}
\newacronym[plural=\gls{mme}s,firstplural=Mobility Management Entities (MMEs)]{mme}{MME}{Mobility Management Entity}
\newacronym{mmtc}{mMTC}{Massive Machine-Type Communications}
\newacronym{mmwave}{mmWave}{millimeter wave}
\newacronym{mpdccp}{MP-DCCP}{Multipath Datagram Congestion Control Protocol}
\newacronym{mps}{MPS}{Multi-Process Service}
\newacronym{mptcp}{MPTCP}{Multipath TCP}
\newacronym{mr}{MR}{Maximum Rate}
\newacronym{mrdc}{MR-DC}{Multi \gls{rat} \gls{dc}}
\newacronym{mse}{MSE}{Mean Square Error}
\newacronym{mss}{MSS}{Maximum Segment Size}
\newacronym{mt}{MT}{Mobile Termination}
\newacronym{mtd}{MTD}{Machine-Type Device}
\newacronym{mtu}{MTU}{Maximum Transmission Unit}
\newacronym{mumimo}{MU-MIMO}{Multi-user \gls{mimo}}
\newacronym{mvno}{MVNO}{Mobile Virtual Network Operator}
\newacronym{nalu}{NALU}{Network Abstraction Layer Unit}
\newacronym{nas}{NAS}{Network Attached Storage}
\newacronym{nbiot}{NB-IoT}{Narrow Band IoT}
\newacronym{nfv}{NFV}{Network Function Virtualization}
\newacronym{nfvi}{NFVI}{Network Function Virtualization Infrastructure}
\newacronym{nic}{NIC}{Network Interface Card}
\newacronym{nlos}{NLOS}{Non-Line-of-Sight}
\newacronym{now}{NOW}{Non Overlapping Window}
\newacronym{nn}{NN}{Neural Network}
\newacronym{nrdz}{NRDZ}{National Radio Dynamic Zone}
\newacronym{nsf}{NSF}{National Science Foundation}
\newacronym{nsm}{NSM}{Network Service Mesh}
\newacronym{nr}{NR}{New Radio}
\newacronym{nrf}{NRF}{Network Repository Function}
\newacronym{nsa}{NSA}{Non Stand Alone}
\newacronym{nse}{NSE}{Network Slicing Engine}
\newacronym{nssf}{NSSF}{Network Slice Selection Function}
\newacronym{ntp}{NTP}{Network Time Protocol}
\newacronym{o2i}{O2I}{Outdoor to Indoor}
\newacronym{oai}{OAI}{OpenAirInterface}
\newacronym{oaic}{OAIC}{Open AI Cellular}
\newacronym{oaicn}{OAI-CN}{\gls{oai} \acrlong{cn}}
\newacronym{oairan}{OAI-RAN}{\acrlong{oai} \acrlong{ran}}
\newacronym{oam}{OAM}{Operations, Administration and Maintenance}
\newacronym[plural=\gls{obu}s,firstplural=Onboard Units (OBUs)]{obu}{OBU}{Onboard Unit}
\newacronym{odc}{ODC}{ORAN Development Company}
\newacronym{ofdm}{OFDM}{Orthogonal Frequency Division Multiplexing}
\newacronym{olia}{OLIA}{Opportunistic Linked Increase Algorithm}
\newacronym{omec}{OMEC}{Open Mobile Evolved Core}
\newacronym{onap}{ONAP}{Open Network Automation Platform}
\newacronym{onf}{ONF}{Open Networking Foundation}
\newacronym{onos}{ONOS}{Open Networking Operating System}
\newacronym{oom}{OOM}{\gls{onap} Operations Manager}
\newacronym{opnfv}{OPNFV}{Open Platform for \gls{nfv}}
\newacronym{orbit}{ORBIT}{Open-Access Research Testbed for Next-Generation Wireless Networks}
\newacronym{os}{OS}{Operating System}
\newacronym{osc}{OSC}{O-RAN Software Community}
\newacronym{osm}{OSM}{Open Street Map}
\newacronym{oss}{OSS}{Operations Support System}
\newacronym{pa}{PA}{Position-aware}
\newacronym{pase}{PASE}{Prioritization, Arbitration, and Self-adjusting Endpoints}
\newacronym{pawr}{PAWR}{Platforms for Advanced Wireless Research}
\newacronym{pbch}{PBCH}{Physical Broadcast Channel}
\newacronym{pci}{PCI}{Peripheral Component Interconnect}
\newacronym{pcef}{PCEF}{Policy and Charging Enforcement Function}
\newacronym{pcfich}{PCFICH}{Physical Control Format Indicator Channel}
\newacronym{pcrf}{PCRF}{Policy and Charging Rules Function}
\newacronym{pdcch}{PDCCH}{Physical Downlink Control Channel}
\newacronym{pdcp}{PDCP}{Packet Data Convergence Protocol}
\newacronym{pdsch}{PDSCH}{Physical Downlink Shared Channel}
\newacronym{pdu}{PDU}{Packet Data Unit}
\newacronym{pdp}{PDP}{Power Delay Profile}
\newacronym{pf}{PF}{Proportional Fair}
\newacronym{pgw}{PGW}{Packet Gateway}
\newacronym{ph}{PH}{Power Headroom}
\newacronym{phich}{PHICH}{Physical Hybrid ARQ Indicator Channel}
\newacronym{phy}{PHY}{Physical}
\newacronym{pl}{PL}{Path Loss}
\newacronym{pmch}{PMCH}{Physical Multicast Channel}
\newacronym{pmi}{PMI}{Precoding Matrix Indicators}
\newacronym{powder}{POWDER}{Platform for Open Wireless Data-driven Experimental Research}
\newacronym{ppo}{PPO}{Proximal Policy Optimization}
\newacronym{ppp}{PPP}{Poisson Point Process}
\newacronym{prach}{PRACH}{Physical Random Access Channel}
\newacronym{prb}{PRB}{Physical Resource Block}
\newacronym{psnr}{PSNR}{Peak Signal to Noise Ratio}
\newacronym{pss}{PSS}{Primary Synchronization Signal}
\newacronym{pucch}{PUCCH}{Physical Uplink Control Channel}
\newacronym{pusch}{PUSCH}{Physical Uplink Shared Channel}
\newacronym{qam}{QAM}{Quadrature Amplitude Modulation}
\newacronym{qci}{QCI}{\gls{qos} Class Identifier}
\newacronym{qoe}{QoE}{Quality of Experience}
\newacronym{qos}{QoS}{Quality of Service}
\newacronym{qtgui}{QT-GUI}{QT Graphical User Interface}
\newacronym{qsfp28}{QSFP28}{Quad Small Form-factor Pluggable 28}
\newacronym{quic}{QUIC}{Quick UDP Internet Connections}
\newacronym{rach}{RACH}{Random Access Channel}
\newacronym{ran}{RAN}{Radio Access Network}
\newacronym[firstplural=Radio Access Technologies (RATs)]{rat}{RAT}{Radio Access Technology}
\newacronym{rb}{RB}{Resource Block}
\newacronym{rcn}{RCN}{Research Coordination Network}
\newacronym{rdma}{RDMA}{Remote Direct Memory Access}
\newacronym{rec}{REC}{Radio Edge Cloud}
\newacronym{red}{RED}{Random Early Detection}
\newacronym{renew}{RENEW}{Reconfigurable Eco-system for Next-generation End-to-end Wireless}
\newacronym{rf}{RF}{Radio Frequency}
\newacronym{rfc}{RFC}{Request for Comments}
\newacronym{rfr}{RFR}{Random Forest Regressor}
\newacronym{ric}{RIC}{RAN Intelligent Controller}
\newacronym{near-rt-ric}{near-RT-RIC}{near-RT-\gls{ric}}
\newacronym{non-rt}{non-RT}{non-Real-Time}
\newacronym{near-rt}{near-RT}{near-Real-Time}
\newacronym{rlc}{RLC}{Radio Link Control}
\newacronym{rlf}{RLF}{Radio Link Failure}
\newacronym{rlnc}{RLNC}{Random Linear Network Coding}
\newacronym{rmse}{RMSE}{Root Mean Squared Error}
\newacronym{rnis}{RNIS}{Radio Network Information Service}
\newacronym{rr}{RR}{Round Robin}
\newacronym{rrc}{RRC}{Radio Resource Control}
\newacronym{rrm}{RRM}{Radio Resource Management}
\newacronym{rru}{RRU}{Remote Radio Unit}
\newacronym{rs}{RS}{Remote Server}
\newacronym{rsrp}{RSRP}{Reference Signal Received Power}
\newacronym{rsrq}{RSRQ}{Reference Signal Received Quality}
\newacronym{rss}{RSS}{Received Signal Strength}
\newacronym{rssi}{RSSI}{Received Signal Strength Indicator}
\newacronym{rsu}{RSU}{Road-Side Unit}
\newacronym{rt}{RT}{Real-Time}
\newacronym{rtt}{RTT}{Round Trip Time}
\newacronym{ru}{RU}{Radio Unit}
\newacronym{rw}{RW}{Receive Window}
\newacronym{rx}{RX}{Receiver}
\newacronym{s1ap}{S1AP}{S1 Application Protocol}
\newacronym{sa}{SA}{standalone}
\newacronym{sack}{SACK}{Selective Acknowledgment}
\newacronym{sap}{SAP}{Service Access Point}
\newacronym{sas}{SAS}{Spectrum Access System}
\newacronym{sc2}{SC2}{Spectrum Collaboration Challenge}
\newacronym{scef}{SCEF}{Service Capability Exposure Function}
\newacronym{sch}{SCH}{Secondary Cell Handover}
\newacronym{scoot}{SCOOT}{Split Cycle Offset Optimization Technique}
\newacronym{sfp+}{SFP+}{Small Form-factor Pluggable Plus}
\newacronym{sctp}{SCTP}{Stream Control Transmission Protocol}
\newacronym{sdap}{SDAP}{Service Data Adaptation Protocol}
\newacronym{sd}{SD}{Standard Deviation}
\newacronym{sdk}{SDK}{Software Development Kit}
\newacronym{sdm}{SDM}{Space Division Multiplexing}
\newacronym{sdma}{SDMA}{Spatial Division Multiple Access}
\newacronym{sdn}{SDN}{Software-defined Networking}
\newacronym{sdr}{SDR}{Software-defined Radio}
\newacronym{seba}{SEBA}{SDN-Enabled Broadband Access}
\newacronym{sgsn}{SGSN}{Serving GPRS Support Node}
\newacronym{sgw}{SGW}{Service Gateway}
\newacronym{shm}{SHM}{Shared Memory}
\newacronym{si}{SI}{Study Item}
\newacronym{sib}{SIB}{Secondary Information Block}
\newacronym{sinr}{SINR}{Signal to Interference plus Noise Ratio}
\newacronym{sip}{SIP}{Session Initiation Protocol}
\newacronym{siso}{SISO}{Single Input, Single Output}
\newacronym{sla}{SLA}{Service Level Agreement}
\newacronym{sm}{SM}{Service Model}
\newacronym{smf}{SMF}{Session Management Function}
\newacronym{smo}{SMO}{Service Management and Orchestration}
\newacronym{sms}{SMS}{Short Message Service}
\newacronym{smsgmsc}{SMS-GMSC}{\gls{sms}-Gateway}
\newacronym{snr}{SNR}{Signal-to-Noise-Ratio}
\newacronym{son}{SON}{Self-Organizing Network}
\newacronym{sptcp}{SPTCP}{Single Path TCP}
\newacronym{srb}{SRB}{Service Radio Bearer}
\newacronym{srn}{SRN}{Standard Radio Node}
\newacronym{srs}{SRS}{Sounding Reference Signal}
\newacronym{ss}{SS}{Synchronization Signal}
\newacronym{sss}{SSS}{Secondary Synchronization Signal}
\newacronym{st}{ST}{Spanning Tree}
\newacronym{svc}{SVC}{Scalable Video Coding}
\newacronym{synce}{SyncE}{Synchronous Ethernet}
\newacronym{tai}{TAI}{International Atomic Time}
\newacronym{tb}{TB}{Transport Block}
\newacronym{tcp}{TCP}{Transmission Control Protocol}
\newacronym{tdd}{TDD}{Time Division Duplexing}
\newacronym{tdm}{TDM}{Time Division Multiplexing}
\newacronym{tdma}{TDMA}{Time Division Multiple Access}
\newacronym{tf}{TF}{Tensorflow}
\newacronym{tfl}{TfL}{Transport for London}
\newacronym{tfrc}{TFRC}{TCP-Friendly Rate Control}
\newacronym{tft}{TFT}{Traffic Flow Template}
\newacronym{tgen}{TGEN}{Traffic Generator}
\newacronym{tip}{TIP}{Telecom Infra Project}
\newacronym{tm}{TM}{Transparent Mode}
\newacronym{to}{TO}{Telco Operator}
\newacronym{toa}{ToA}{Time of Arrival}
\newacronym{tl}{TL}{Transfer Learning}
\newacronym{tr}{TR}{Technical Report}
\newacronym{trt}{TRT}{TensorRT}
\newacronym{trp}{TRP}{Transmitter Receiver Pair}
\newacronym{ts}{TS}{Technical Specification}
\newacronym{tti}{TTI}{Transmission Time Interval}
\newacronym{ttt}{TTT}{Time-to-Trigger}
\newacronym{tx}{TX}{Transmitter}
\newacronym{uas}{UAS}{Unmanned Aerial System}
\newacronym{uav}{UAV}{Unmanned Aerial Vehicle}
\newacronym{udm}{UDM}{Unified Data Management}
\newacronym{udp}{UDP}{User Datagram Protocol}
\newacronym{udr}{UDR}{Unified Data Repository}
\newacronym{ue}{UE}{User Equipment}
\newacronym{uhd}{UHD}{\gls{usrp} Hardware Driver}
\newacronym{ul}{UL}{uplink}
\newacronym{um}{UM}{Unacknowledged Mode}
\newacronym{uml}{UML}{Unified Modeling Language}
\newacronym{upa}{UPA}{Uniform Planar Array}
\newacronym{upf}{UPF}{User Plane Function}
\newacronym{urllc}{URLLC}{Ultra Reliable and Low Latency Communication}
\newacronym{usa}{U.S.}{United States}
\newacronym{usim}{USIM}{Universal Subscriber Identity Module}
\newacronym{usrp}{USRP}{Universal Software Radio Peripheral}
\newacronym{utc}{UTC}{Coordinated Universal Time}
\newacronym{vim}{VIM}{Virtualization Infrastructure Manager}
\newacronym{vlan}{VLAN}{Virtual Local Area Network}
\newacronym{vm}{VM}{Virtual Machine}
\newacronym{vnf}{VNF}{Virtual Network Function}
\newacronym{volte}{VoLTE}{Voice over \gls{lte}}
\newacronym{voltha}{VOLTHA}{Virtual OLT HArdware Abstraction}
\newacronym{vr}{VR}{Virtual Reality}
\newacronym{vran}{vRAN}{Virtualized \gls{ran}}
\newacronym{vss}{VSS}{Video Streaming Server}
\newacronym{wbf}{WBF}{Wired Bias Function}
\newacronym{wf}{WF}{Wired-first}
\newacronym{wi}{WI}{Wireless InSite}
\newacronym{wlan}{WLAN}{Wireless Local Area Network}
\newacronym{zmq}{ZMQ}{ZeroMQ}
\newacronym{pnf}{PNF}{Physical Network Function}
\newacronym{drl}{DRL}{Deep Reinforcement Learning}
\newacronym{mtc}{MTC}{Machine-type Communications}
\newacronym{v2x}{V2X}{Vehicle-to-everything}
\newacronym{cast}{\textit{CaST}}{Channel emulation generator and Sounder Toolchain}
\newacronym{abr}{ABR}{Adaptive Bitrate Streaming}
\newacronym{arc}{ARC}{Aerial RAN CoLab}
\newacronym{dsp}{DSP}{Digital Signal Processing}
\newacronym{ota}{OTA}{Over-the-Air}
\newacronym{bom}{BoM}{Bill of Materials}
\newacronym{frand}{FRAND}{Fair, Reasonable, And Non-Discriminatory}
\newacronym{nvipc}{NVIPC}{NVIDIA Inter-Process Communication}
\newacronym{hdr}{HDR}{High Dynamic Range}
\newacronym{ipc}{IPC}{Inter-Process Communication}
\newacronym{uci}{UCI}{Uplink Control Indication}
\newacronym{cbrs}{CBRS}{Citizen Broadband Radio Service}
\newacronym{ptp}{PTP}{Precision Timing Protocol}
\newacronym{scf}{SCF}{Small Cell Forum}
\newacronym{re}{RE}{Resource Element}
\newacronym{fp16}{FP16}{Float16}
\newacronym{fp32}{FP32}{Float32}
\newacronym{int32}{INT32}{32-bit Integer}
\newacronym{tdl}{TDL}{Tapped Delay Line}
\newacronym{fh}{FH}{Front-haul}
\newacronym{ta}{TA}{Timing Advance}
\newacronym{cfo}{CFO}{Carrier Frequency Offset}
\newacronym{sir}{SIR}{Signal to Interference Ratio}
\newacronym{mmse}{MMSE}{Minimum Mean Square Error}
\newacronym{irc}{IRC}{Interference Rejection Combining}
\newacronym{llm}{LLM}{Large Language Model}
\newacronym{tco}{TCO}{Total Cost of Ownership}
\newacronym{fec}{FEC}{Forward Error Correction}
\newacronym{oran}{O-RAN}{Open Radio Access Network}
\newacronym{airan}{AI-RAN}{Artificial Intelligence Radio Access Network}
\newacronym{ssb}{SSB}{Synchronization Signal Block}
\newacronym{fp8}{FP8}{8-bit Floating Point}
\newacronym{capex}{CapEx}{Capital Expenditure}
\newacronym{opex}{OpEx}{Operational Expenditure}
\newacronym{cagr}{CAGR}{Compound Annual Growth Rate}

%% file: server_table.tex
\begin{table*}[t]
  \small
  \centering
  \caption{Summary of Server Characteristics, Baseband Capacity, and Economic Efficiency}
  \label{tab:capacity_summary}
  \begin{tabular}{l l l r r r r r r r r r l}
    \toprule
    \textbf{Platform} &
    \textbf{Accel.} &
    \textbf{Layer 1} &
    \shortstack{\textbf{Cost}\\\textbf{(\$)}} &
    \shortstack{\textbf{Power}\\\textbf{(W)}} &
    \shortstack{\textbf{L}\\ \textbf{DL/UL}} &
    \shortstack{$\mathbf{N_C}$\\~} &
    \shortstack{\textbf{BW}\\\textbf{(MHz)}} &
    \shortstack{\textbf{B}\\\textbf{(MHz)}} &
    \shortstack{$\mathbf{\eta_C}$\\\textbf{(MHz/\$)}} &
    \shortstack{$\mathbf{\eta_O}$\\\textbf{(MHz/W)}} &
    \textbf{Ref.} \\
    \midrule
    ARS-111GL    & GH200  & Aerial  & 45000 & 1200 & 4/4 & 40 & 100  & 16000 & 0.36 & 13.33 & \cite{nvidia2023enabling} \\
    ARS-111GL    & GH200  & Aerial  & 45000 & 1200 & 16/8 & 6 & 100  & 9600 & 0.21 & 8.00 & \cite{aerial25-2spec} \\
    \midrule
    \shortstack{mixed 3:1\\w. avg. }& GH200  & Aerial  & 45000 & 1200 & -- & -- & -- & 14400 &   0.32 & 12.00 & \\
    \midrule
    EGX74I       & VRB1   & FlexRAN & 6000 & 300 & 16/8 & 6 & 100  & 9600 & 1.60 & 32.00 & \cite{qctvrb100} \\
    EGX74I       & VRB1   & FlexRAN & 6000 & 300 & 4/4 & 36 & 10  & 1440 & 0.24 & 4.80 & \cite{qctvrb100} \\
    DL110        & VRB1   & FlexRAN & 7200 & 300 & 4/4 & 18 & 20  & 1440 & 0.20 & 4.80 & \cite{intelvrandl110} \\
    DL110        & VRB1   & FlexRAN & 7200 & 300 & 16/8 & 6 & 100  & 9600 & 1.33 & 32.00 & \cite{intelvrandl110} \\
    \midrule
    \shortstack{mixed 3:1\\w. avg. }& VRB1   & FlexRAN & 6600 & 300 & -- & -- & -- & 3480 &   0.70 & 11.60 & \\
   \bottomrule
  \end{tabular}
\end{table*}

%% file: figures/tco_pgfplot.tex
% Reset stacking between bar groups so we get 2 separate stacked bars per x
\makeatletter
\newcommand\resetstackedplots{%
  \pgfplots@stacked@isfirstplottrue
  \addplot [forget plot, draw=none] coordinates {(Micro,0) (Macro,0) (Mixed,0)};
}
\makeatother

\begin{tikzpicture}
\begin{axis}[
  ybar stacked,
  ymin=0, ymax=9000,
  width=0.95\linewidth,
  height=0.45\linewidth,
  ylabel={Cost (\$)},
  x tick label style={font=\footnotesize},
  y tick label style={font=\footnotesize},
  label style={font=\footnotesize},
  legend style={font=\footnotesize, at={(0.5,1.02)}, anchor=south, legend columns=2, column sep=0.2cm},
  symbolic x coords={Micro, Macro, Mixed},
  xtick=data,
  bar width=20pt,
  enlarge x limits=0.2,
  yticklabel={\pgfmathparse{\tick/1000}\pgfmathprintnumber{\pgfmathresult}k},
  grid=major,
  grid style={dashed,gray!30},
]

  % Aerial stack (red): bar shift left
  \addplot+[forget plot, fill=red, draw=lightgray, line width=0.5pt, bar shift=-10pt]
    table[x=cell_type, y=aerial_capex, col sep=comma] {data/tco.csv};
  \addplot+[forget plot, fill=red, pattern=north east lines, pattern color=red!80!black,
    draw=lightgray, line width=0.5pt, bar shift=-10pt]
    table[x=cell_type, y=aerial_opex, col sep=comma] {data/tco.csv};

  \resetstackedplots

  % FlexRAN stack (blue): bar shift right
  \addplot+[forget plot, fill=blue, draw=lightgray, line width=0.5pt, bar shift=10pt]
    table[x=cell_type, y=flexran_capex, col sep=comma] {data/tco.csv};
  \addplot+[forget plot, fill=blue, pattern=north east lines, pattern color=blue!80!black,
    draw=lightgray, line width=0.5pt, bar shift=10pt]
    table[x=cell_type, y=flexran_opex, col sep=comma] {data/tco.csv};

  % Legend: Red Aerial, Blue FlexRAN, Solid CapEx, Dashed OpEx
  \addlegendimage{fill=red, area legend}
  \addlegendentry{Aerial}
  \addlegendimage{fill=blue, area legend}
  \addlegendentry{FlexRAN}
  \addlegendimage{fill=gray, area legend}
  \addlegendentry{CapEx}
  \addlegendimage{fill=gray, pattern=north east lines, area legend}
  \addlegendentry{OpEx (10y)}

\end{axis}
\end{tikzpicture}

%% file: figures/daily_usage_pgfplot.tex
\begin{tikzpicture}
\begin{axis}[
    width=0.48\textwidth,
    height=0.25\textwidth,
    xlabel={h},
    ylabel={Normalized Demand},
    xmin=0, xmax=167,
    ymin=0, ymax=1.1,
    xtick={0,24,48,72,96,120,144,168},
    legend pos=north east,
    grid=major,
    grid style={dashed,gray!30},
    legend style={font=\footnotesize},
    tick label style={font=\footnotesize},
    label style={font=\footnotesize},
    every axis plot/.append style={thick},
]

% RAN demand
\addplot[
    color=red,
    mark size=2pt,
] table[x=time, y=call_in, col sep=comma] {data/weekly_call_in.csv};
\addlegendentry{$\lambda_\mathrm{RAN}(h)$}

% LLM demand
\addplot[
    color=blue!70!black,
    mark size=2pt,
] table[x=time, y=llm, col sep=comma] {data/llm_trend.csv};
\addlegendentry{$\lambda_\mathrm{LLM}(h)$}

\end{axis}
\end{tikzpicture}

%% file: figures/gpu_usage_pgfplot.tex
\begin{tikzpicture}
\begin{groupplot}[
  group style={
    group size=1 by 2,
    vertical sep=0.15cm,
  },
  width=0.95\linewidth,
  height=0.40\linewidth,
  xlabel={h},
  xmin=0, xmax=168,
  xtick={0,24,48,72,96,120,144,168},
  stack plots=y,
  area style,
  axis on top=false,
  grid=major,
  grid style={dashed,black!30},
  legend style={font=\footnotesize, at={(0.5,1.02)}, anchor=south, legend columns=3, column sep=0.1cm},
  tick label style={font=\footnotesize},
  label style={font=\footnotesize},
]

% Top: no x-axis labels
\nextgroupplot[ylabel={GPUs}, ymin=0, xticklabels={}, xlabel={}]
  \addplot[fill=blue!80!white, fill opacity=0.7, draw=none]
    table[x=hour, y=ran, col sep=comma] {data/gpu_usage_cagr1.csv} \closedcycle;
  \addplot[fill=orange!80!white, fill opacity=0.7, draw=none]
    table[x=hour, y=llm, col sep=comma] {data/gpu_usage_cagr1.csv} \closedcycle;
  \addplot[fill=gray!30!white, fill opacity=0.7, draw=none]
    table[x=hour, y=idle, col sep=comma] {data/gpu_usage_cagr1.csv} \closedcycle;
  \addlegendentry{$G^\mathrm{RAN}(0,h)$}
  \addlegendentry{$G_{alloc}^\mathrm{LLM}(0,h)$}
  \addlegendentry{Idle}

% Bottom: full x-axis
\nextgroupplot[ylabel={GPUs}, ymin=0]
  \addplot[fill=blue!80!white, fill opacity=0.7, draw=none]
    table[x=hour, y=ran, col sep=comma] {data/gpu_usage_cagr12.csv} \closedcycle;
  \addplot[fill=orange!80!white, fill opacity=0.7, draw=none]
    table[x=hour, y=llm, col sep=comma] {data/gpu_usage_cagr12.csv} \closedcycle;
  \addplot[fill=gray!30!white, fill opacity=0.7, draw=none]
    table[x=hour, y=idle, col sep=comma] {data/gpu_usage_cagr12.csv} \closedcycle;

\end{groupplot}
\end{tikzpicture}

%% file: figures/gpu_allocation_pgfplot.tex
\begin{tikzpicture}
\begin{groupplot}[
  group style={
    group size=1 by 2,
    vertical sep=0.15cm,
  },
  width=0.95\linewidth,
  height=0.45\linewidth,
  xlabel={w [weeks]},
  xmin=0, xmax=520,
  axis on top=false,
  grid=major,
  grid style={dashed,gray!30},
  legend style={font=\footnotesize, at={(0.45,1.04)}, anchor=south, legend columns=2, transpose legend, column sep=0.1cm},
  tick label style={font=\footnotesize},
  label style={font=\footnotesize},
]

% Top: no x-axis labels
% Blue = RAN, Orange = LLM; line style = density parameter
\nextgroupplot[ylabel={GPUs}, xticklabels={}, xlabel={}, ymin=0]
  \addplot[color=blue, very thick, solid, mark=none]
    table[x expr=\thisrow{year}*52, y=ran, col sep=comma] {data/gpu_allocation_w0.csv};
  \addplot[color=orange!90!yellow, very thick, dashed, mark=none]
    table[x expr=\thisrow{year}*52, y=llm_3x, col sep=comma] {data/gpu_allocation_w0.csv};
  % \addplot[color=orange!90!yellow, very thick, dotted, mark=none, forget plot]
  %   table[x expr=\thisrow{year}*52, y=llm_10x, col sep=comma] {data/gpu_allocation_w0.csv};
  \addplot[color=orange!90!yellow, very thick, dashdotted, mark=none, forget plot]
    table[x expr=\thisrow{year}*52, y=llm_article, col sep=comma] {data/gpu_allocation_w0.csv};
  \addplot[color=black, very thick, solid, mark=none, forget plot]
    coordinates {(0,35) (520,35)};
  \addlegendentry{$\hat{G}^\mathrm{RAN}(w)$}
  \addlegendentry{$\hat{G}^\mathrm{LLM}_\mathrm{alloc}(w)$}
  \addlegendimage{line legend, dashed, draw=black!70, very thick}
  \addlegendentry{$\rho_\mathrm{dens}=3$}
  \addlegendimage{line legend, dashdotted, draw=black!70, very thick}
  \addlegendentry{$\rho_\mathrm{dens}=12.87$}
  \addlegendimage{line legend, solid, draw=black,  thick}
  \addlegendentry{$G^\mathrm{tot}$}

  \vspace{1em}
% Bottom: full x-axis
\nextgroupplot[ylabel={GPUs}, ymin=0, ymax=225, ytick={0,50,100,150}]
  \addplot[color=blue, very thick, solid, mark=none, forget plot]
    table[x expr=\thisrow{year}*52, y=ran, col sep=comma] {data/gpu_allocation_w10.csv};
  \addplot[color=orange!90!yellow, very thick, dashed, mark=none, forget plot]
    table[x expr=\thisrow{year}*52, y=llm_3x, col sep=comma] {data/gpu_allocation_w10.csv};
  % \addplot[color=orange!90!yellow, very thick, dotted, mark=none, forget plot]
  %   table[x expr=\thisrow{year}*52, y=llm_10x, col sep=comma] {data/gpu_allocation_w10.csv};
  \addplot[color=orange!90!yellow, very thick, dashdotted, mark=none, forget plot]
    table[x expr=\thisrow{year}*52, y=llm_article, col sep=comma] {data/gpu_allocation_w10.csv};
  \addplot[color=black, thick, solid, mark=none, forget plot]
    coordinates {(0,215) (520,215)};

\end{groupplot}
\end{tikzpicture}

%% file: figures/llm_revenue_pgfplot.tex
\begin{tikzpicture}
\begin{groupplot}[
  group style={
    group size=1 by 2,
    vertical sep=0.15cm,
  },
  width=0.95\linewidth,
  height=0.45\linewidth,
  xlabel={Week},
  xmin=0, xmax=520,
  stack plots=false,
  axis on top=false,
  grid=major,
  grid style={solid,black!30},
  legend style={font=\footnotesize, at={(0.5,1.04)}, anchor=south, legend columns=4},
  tick label style={font=\footnotesize},
  label style={font=\footnotesize},
  y filter/.code={\pgfmathparse{\pgfmathresult/1000}},
]

\nextgroupplot[ylabel={$R_\mathrm{LLM}(w)$ ~[k\$]}, ymin=0, xticklabels={}, xlabel={}]
  \addplot[blue, very thick] table[x=week, y=rev_k1, col sep=comma] {data/llm_revenue_w0.csv};
  \addplot[red, very thick] table[x=week, y=rev_k1_25, col sep=comma] {data/llm_revenue_w0.csv};
  \addplot[green!60!black, very thick] table[x=week, y=rev_k1_5, col sep=comma] {data/llm_revenue_w0.csv};
  \addplot[orange, very thick] table[x=week, y=rev_k2, col sep=comma] {data/llm_revenue_w0.csv};
  \addlegendentry{$k{=}1$}
  \addlegendentry{$k{=}1.25$}
  \addlegendentry{$k{=}1.5$}
  \addlegendentry{$k{=}2$}

\nextgroupplot[ylabel={$R_\mathrm{LLM}(w)$ ~[k\$]}, ymin=0]
  \addplot[blue, very thick] table[x=week, y=rev_k1, col sep=comma] {data/llm_revenue_w10.csv};
  \addplot[red, very thick] table[x=week, y=rev_k1_25, col sep=comma] {data/llm_revenue_w10.csv};
  \addplot[green!60!black, very thick] table[x=week, y=rev_k1_5, col sep=comma] {data/llm_revenue_w10.csv};
  \addplot[orange, very thick] table[x=week, y=rev_k2, col sep=comma] {data/llm_revenue_w10.csv};

\end{groupplot}
\end{tikzpicture}

%% file: figures/milan_tco_pgfplot.tex
\makeatletter
\newcommand\resetstackedplots{%
  \pgfplots@stacked@isfirstplottrue
  \addplot [forget plot, draw=none] coordinates {(Scenario 1,0) (Scenario 2,0)};
}
\makeatother

\begin{tikzpicture}
\begin{axis}[
  ybar stacked,
  ymin=0,
  width=0.80\linewidth,
  height=0.45\linewidth,
  ylabel={C ~[M\$]},
  x tick label style={font=\footnotesize},
  y tick label style={font=\footnotesize},
  label style={font=\footnotesize},
  legend style={font=\footnotesize, at={(1.02,0.5)}, anchor=west, legend columns=1},
  symbolic x coords={Scenario 1, Scenario 2},
  xtick=data,
  bar width=20pt,
  enlarge x limits=0.5,
  grid=major,
  grid style={dashed,gray!30},
]

  \addplot+[forget plot, fill=red, draw=lightgray, line width=0.5pt, bar shift=-10pt]
    table[x=scenario, y expr=\thisrow{aerial_capex}/1e6, col sep=comma] {data/milan_tco.csv};
  % \addplot+[forget plot, fill=red, pattern=north east lines, pattern color=red!80!black,
  %   draw=lightgray, line width=0.5pt, bar shift=-10pt]
  %   table[x=scenario, y expr=\thisrow{aerial_opex}/1e6, col sep=comma] {data/milan_tco.csv};

  \resetstackedplots

  \addplot+[forget plot, fill=blue, draw=lightgray, line width=0.5pt, bar shift=10pt]
    table[x=scenario, y expr=\thisrow{flexran_capex}/1e6, col sep=comma] {data/milan_tco.csv};
  % \addplot+[forget plot, fill=blue, pattern=north east lines, pattern color=blue!80!black,
  %   draw=lightgray, line width=0.5pt, bar shift=10pt]
  %   table[x=scenario, y expr=\thisrow{flexran_opex}/1e6, col sep=comma] {data/milan_tco.csv};

  \addlegendimage{fill=red, area legend}
  \addlegendentry{Aerial}
  \addlegendimage{fill=blue, area legend}
  \addlegendentry{FlexRAN}
  % \addlegendimage{fill=gray, area legend}
  % \addlegendentry{CapEx}
  % \addlegendimage{fill=gray, pattern=north east lines, area legend}
  % \addlegendentry{OpEx}

\end{axis}
\end{tikzpicture}

%% file: figures/llm_revenue_cumulative_pgfplot.tex
\begin{tikzpicture}
\begin{groupplot}[
  group style={
    group size=1 by 2,
    vertical sep=0.15cm,
  },
  width=0.95\linewidth,
  height=0.45\linewidth,
  xlabel={w [weeks]},
  xmin=0, xmax=520,
  ylabel={R(w) [M\$]},
  ymin=0,
  yticklabel={\pgfmathprintnumber[fixed,precision=1]{\tick}},
  grid=major,
  grid style={solid,black!30},
  legend style={font=\footnotesize, at={(0.5,1.02)}, anchor=south, legend columns=4},
  tick label style={font=\footnotesize},
  label style={font=\footnotesize},
  %restrict y to domain=0:11
]

\nextgroupplot[xticklabels={}, xlabel={}]
  \addplot[blue, very thick] table[x=week, y expr=\thisrow{rev_k1}/1000000, col sep=comma] {data/llm_cumulative_w0.csv};
  \addplot[red, very thick] table[x=week, y expr=\thisrow{rev_k1_25}/1000000, col sep=comma] {data/llm_cumulative_w0.csv};
  \addplot[green!60!black, very thick] table[x=week, y expr=\thisrow{rev_k1_5}/1000000, col sep=comma] {data/llm_cumulative_w0.csv};
  \addplot[orange, very thick] table[x=week, y expr=\thisrow{rev_k2}/1000000, col sep=comma] {data/llm_cumulative_w0.csv};
  \addplot[dashed, black, very thick] table[x=week, y expr=\thisrow{ref}/1000000, col sep=comma] {data/tco_ref_w0.csv};
  \addlegendentry{$k{=}1$}
  \addlegendentry{$k{=}1.25$}
  \addlegendentry{$k{=}1.5$}
  \addlegendentry{$k{=}2$}
  \addlegendentry{Investment}

\nextgroupplot
  \addplot[blue, very thick] table[x=week, y expr=\thisrow{rev_k1}/1000000, col sep=comma] {data/llm_cumulative_wW.csv};
  \addplot[red, very thick] table[x=week, y expr=\thisrow{rev_k1_25}/1000000, col sep=comma] {data/llm_cumulative_wW.csv};
  \addplot[green!60!black, very thick] table[x=week, y expr=\thisrow{rev_k1_5}/1000000, col sep=comma] {data/llm_cumulative_wW.csv};
  \addplot[orange, very thick] table[x=week, y expr=\thisrow{rev_k2}/1000000, col sep=comma] {data/llm_cumulative_wW.csv};
  \addplot[dashed, black, very thick] table[x=week, y expr=\thisrow{ref}/1000000, col sep=comma] {data/tco_ref_wW.csv};

\end{groupplot}
\end{tikzpicture}

%% file: figures/roi_pgfplot.tex
\pgfplotsset{compat=1.18}
\begin{tikzpicture}
\begin{axis}[
  ybar,
  bar width=12pt,
  width=0.95\linewidth,
  height=0.50\linewidth,
  ylabel={$R/I$},
  xlabel={$k$},
  symbolic x coords={1.0, 1.25, 1.5, 2.0},
  xtick=data,
  ymin=0,
  enlarge x limits=0.2,
  tick label style={font=\footnotesize},
  y tick label style={font=\footnotesize},
  x tick label style={font=\footnotesize},
  legend style={font=\footnotesize, at={(0.5,1.02)}, anchor=south, legend columns=2, column sep=0.2cm},
  grid=major,
  grid style={dashed,gray!30},
]

  \addplot[single ybar legend, fill=blue!60, draw=black!50] table[x=k, y=RI, col sep=comma] {data/roi_w0.csv};
  \addplot[single ybar legend, fill=orange!60, draw=black!50] table[x=k, y=RI, col sep=comma] {data/roi_wW.csv};
  \legend{Scenario 1, Scenario 2}

\end{axis}
\end{tikzpicture}